\documentclass[11pt,dvips]{article}
\usepackage{amssymb}
\usepackage{epsfig}
\usepackage{subfigure}

\def\R{\ensuremath{\mathbb R}}
\def\C{\ensuremath{\mathbb C}}
\def\I{\ensuremath{\mathbb I}}
\def\diff{\ensuremath{\mbox{d}}}

\begin{document}

\begin{titlepage}
\begin{flushright}
MAV-UPC-02/96
\end{flushright}
\vskip1cm
\begin{center}
\textbf{\Large Extended time-delay autosynchronization for the \textit{buck} converter }\\ 
\vskip5mm
\textbf{\large C. Batlle} (carles@mat.upc.es), \textbf{\large E. Fossas} (fossas@mat.upc.es) 
\\ and \textbf{\large G. Olivar} (gerard@mat.upc.es)\\
Departament de Matem\`atica Aplicada i Telem\`atica\\
Universitat Polit\`ecnica de Catalunya\\
M\`odul C3 Campus Nord\\
Gran Capit\`a s/n\\
08034 BARCELONA (Spain)
\vskip1cm
July 1996
\end{center}
\vskip2cm

\begin{center}\textbf{ABSTRACT}\end{center}

Time-delay autosynchronization (TDAS) 
can be used to stabilize unstable periodic orbits in dynamical systems.
The technique involves continuous feedback of signals delayed by the orbit's period. 
One variant, ETDAS, uses information further in the past.
In both cases, the feedback signal vanishes on the target periodic orbit and
hence the stabilized periodic orbit is one of the original dynamical system. 
Furthermore, this control method only requires the knowledge of the period of the unstable orbit. 

In general, the amount of feedback gain needed to achieve stabilization varies with the 
bifurcation parameter(s) of the system, resulting in a domain of control 
which can be computed without having to deal with the explicit integration of time-delay equations.

In this paper we compute the domain of control of the unstable periodic orbits 
of the \textit{buck} converter. The simplicity of the nonlinearity of this converter 
allows us to obtain a closed analytical expression for the curve $g:S^1 \rightarrow\C$ 
whose index determines the stability. We perform detailed numerical explorations 
of this index for both period-1 and period-2 orbits for several configurations of the feedback
scheme. We run several simulations of the controlled system and discuss the results.
\end{titlepage}

\section{Introduction}
\label{intro}

Control of chaos, meaning supression of the chaotic regime in a system by means of a small, time-dependent perturbation,
has been a subject of interest in recent years \cite{SGOY}. In \cite{OGY}, it was pointed out that the many unstable periodic orbits
(UPOs) embedded in a strange attractor could be used to produce regular behaviour to the advantage of engineers trying to control
nonlinear systems in which chaotic fluctuations are present but undesirable. It has also been speculated that UPOs may play
a key role in the regulation of complex biological systems.

Two main groups of methods of control of chaos such that the feedback perturbation vanishes on the target orbit have been considered
in the literature:

\begin{itemize}
\item The method proposed by Ott, Grebogi and Yorke \cite{OGY}, where small perturbations to an accessible parameter are introduced.
 The method exploits the fact that during its wandering over the strange attractor, the system will eventually come near the target
 UPO on a given Poincar\'e section. When this happens, and only then, a small perturbation is applied to the parameter so as to make
 the orbit land on the stable variety of the target orbit the next time it crosses the Poincar\'e section. As a drawback, the method
 is not suitable for complex systems, since a nontrivial computer analysis must be performed at each crossing of the Poincar\'e section.
 Also, small noise can drive the orbit away from the target orbit, and the control method must then wait for a while until the
 system comes near to the target orbit again.
\item The method proposed by Pyragas \cite{TDAS}, called time-delayed autosynchronization (TDAS), involves a control signal formed with 
the difference between the current state of the system and the state of the system delayed by one period of the UPO. One variant, ETDAS,
proposed by Socolar \textit{et al} \cite{ETDAS}, uses a particular linear combination of signals from the system delayed by integer
multiples of the UPO's period. Still another variant, proposed by de Sousa Vieira \textit{et al} \cite{dSV} uses a nonlinear
function of the difference between the present state and the delayed state. TDAS and its variants have the advantage that the only
information needed about the target orbit is its period and that no computer processing must be done to generate the control
signal. The method has even been applied to systems described by partial differential equations \cite{BS2}.
 In general, the feedback gain which succesfully stabilizes the orbit lies in a finite, and often narrow, orbit-dependent range.
In the space of the feedback gain and the bifurcation parameter(s) of the system, the region where the TDAS can be applied with success
is called the domain of control. In \cite{BS} a method was proposed to compute the domain of control of a given system without having to
explicitly integrate the resulting time-delay equations, which is a nontrivial matter due to the choice of initial conditions \cite{Hale}.
Essentially, the method reduces to the computation of the index around the origin of a curve in the complex plane.
\end{itemize}

In this paper we address the problem of stabilizing the UPOs of the \textit{buck} converter by means of ETDAS. The chaotic behaviour
of this converter and the computation of its UPOs have been extensively studied in \cite{DH}, \cite{diB} and \cite{FO}, and methods of control based
in OGY techniques and others have been analyzed \cite{CB}\cite{BFO}. The main result of our work is that the function $g$ 
fron the unit circle 
to the complex plane whose index determines the success of ETDAS can be analytically computed for the \textit{buck} converter. The index
can then be numerically evaluated and the domain of control can be easily constructed.

The paper is organized as follows. In Section \ref{buck} we present the three schemes that we have studied to feed back the control
signal to the buck converter. In Section \ref{d_of_c} we review the method os Bleich and Socolar to compute the domain of control 
and specialize it to variable structure systems. In Section \ref{buck_dc} we analytically compute $g$ 
for the period-$1$ orbits of the \textit{buck}
converter and the three proposed schemes. In Section \ref{buck_num} we perform numerical explorations of the exact $g$ functions obtained
in Section \ref{buck_dc} and draw the corresponding domains of control. In Section \ref{buck_higher} we partially extend our results
to higher period orbits of the \textit{buck} converter and treat in some detail the $2$-periodic orbits. In Section \ref{buck_sim}
we perform explicit simulations of the ETDAS controlled \textit{buck} and get numerical, independent confirmation of our analytical
results, and at the same time illustrate the effectiveness of the ETDAS technique. 
Finally, we state our conclusions in Section \ref{conc}.

\section{Time-delayed feedback for the \textit{buck} converter}
\label{buck}

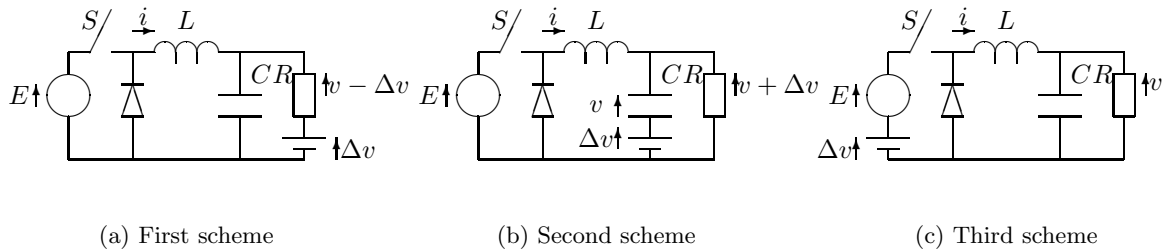
\begin{figure}
\setlength{\unitlength}{2.85mm}
\centering
\subfigure[First scheme]{
\begin{picture}(17,8)
{\small
\put(0,0){\makebox(17,8){}}   
\put(3,1){\line(1,0){11}}     
\put(5.5,3){\line(1,0){1}}     
\put(10,3){\line(1,0){2}}     
\put(13.5,3){\line(1,0){1}}     
\put(5.5,4.5){\line(1,0){1}}     
\put(10,4){\line(1,0){2}}     
\put(13.5,5){\line(1,0){1}}     
\put(3,6){\line(1,0){1}}     
\put(5,6){\line(1,0){2}}     
\put(10,6){\line(1,0){4}}     
\put(3,5){\line(0,1){1}}     
\put(6,1){\line(0,1){2.0}}     
\put(6,4.5){\line(0,1){1.5}}     
\put(8,5.5){\line(0,1){0.5}}     
\put(9,5.5){\line(0,1){0.5}}     
\put(11,4){\line(0,1){2}}     
\put(14,5){\line(0,1){1}}     
\put(3,1){\line(0,1){2}}     
\put(11,1){\line(0,1){2}}     
\put(14,2){\line(0,1){1}}     
\put(13,2){\line(1,0){2}}     
\put(13.5,1.5){\line(1,0){1}}     
\put(14,1){\line(0,1){0.5}}     
\put(4,6){\line(1,1){1}}     
\put(13.5,3){\line(0,1){2}}     
\put(14.5,3){\line(0,1){2}}     
\put(5.5,3){\line(1,3){0.5}}     
\put(6.5,3){\line(-1,3){0.5}}     
\put(7.5,6){\oval(1,1)[t]}       
\put(8.5,6){\oval(1,1)[t]}       
\put(9.5,6){\oval(1,1)[t]}      
\put(3,4){\oval(2,2)}                      
\put(4,6){\line(1,2){1}}     
\put(4,7.5){\makebox(0,0){$S$}}
\put(8.5,7.5){\makebox(0,0){$L$}}
\put(13,5){\makebox(0,0){$R$}}
\put(0.7,4){\makebox(0,0){$E$}}
\put(12,5){\makebox(0,0){$C$}}
\put(6.5,7.7){\makebox(0,0){$i$}}
\put(17.0,4.5){\makebox(0,0){$v-\Delta v$}}
\put(16.5,1.5){\makebox(0,0){$\Delta v$}}
\put(6,7){\vector(1,0){1}}     
\put(1.5,3.5){\vector(0,1){1}}     
\put(15,4){\vector(0,1){1}}     
\put(15.5,1){\vector(0,1){1}}     
}
\end{picture}
}
\ \
\subfigure[Second scheme]{
\begin{picture}(17,8)
{\small
\put(0,0){\makebox(17,8){}}   
\put(3,1){\line(1,0){11}}     
\put(5.5,3){\line(1,0){1}}     
\put(10,3){\line(1,0){2}}     
\put(13.5,3){\line(1,0){1}}     
\put(5.5,4.5){\line(1,0){1}}     
\put(10,4){\line(1,0){2}}     
\put(13.5,5){\line(1,0){1}}     
\put(3,6){\line(1,0){1}}     
\put(5,6){\line(1,0){2}}     
\put(10,6){\line(1,0){4}}     
\put(3,5){\line(0,1){1}}     
\put(6,1){\line(0,1){2.0}}     
\put(6,4.5){\line(0,1){1.5}}     
\put(3,1){\line(0,1){2}}     
\put(8,5.5){\line(0,1){0.5}}     
\put(9,5.5){\line(0,1){0.5}}     
\put(11,4){\line(0,1){2}}     
\put(14,5){\line(0,1){1}}     
\put(11,1){\line(0,1){0.5}}     
\put(11,2){\line(0,1){1}}     
\put(10,2){\line(1,0){2}}     
\put(14,1){\line(0,1){2}}     
\put(10.5,1.5){\line(1,0){1}}     
\put(4,6){\line(1,1){1}}     
\put(13.5,3){\line(0,1){2}}     
\put(14.5,3){\line(0,1){2}}     
\put(5.5,3){\line(1,3){0.5}}     
\put(6.5,3){\line(-1,3){0.5}}     
\put(7.5,6){\oval(1,1)[t]}       
\put(8.5,6){\oval(1,1)[t]}       
\put(9.5,6){\oval(1,1)[t]}      
\put(3,4){\oval(2,2)}                      
\put(4,6){\line(1,2){1}}     
\put(4,7.5){\makebox(0,0){$S$}}
\put(8.5,7.5){\makebox(0,0){$L$}}
\put(13,5){\makebox(0,0){$R$}}
\put(0.7,4){\makebox(0,0){$E$}}
\put(12,5){\makebox(0,0){$C$}}
\put(6.5,7.7){\makebox(0,0){$i$}}
\put(17.0,4.5){\makebox(0,0){$v+\Delta v$}}
\put(8.5,2){\makebox(0,0){$\Delta v$}}
\put(8.5,3.5){\makebox(0,0){$v$}}
\put(6,7){\vector(1,0){1}}     
\put(1.5,3.5){\vector(0,1){1}}     
\put(15,4){\vector(0,1){1}}     
\put(9.5,3){\vector(0,1){1}}     
\put(9.5,1.5){\vector(0,1){1}}     
}
\end{picture}
}
\ \
\subfigure[Third scheme]{
\begin{picture}(17,8)
{\small
\put(0,0){\makebox(17,8){}}   
\put(3,1){\line(1,0){11}}     
\put(5.5,3){\line(1,0){1}}     
\put(10,3){\line(1,0){2}}     
\put(13.5,3){\line(1,0){1}}     
\put(5.5,4.5){\line(1,0){1}}     
\put(10,4){\line(1,0){2}}     
\put(13.5,5){\line(1,0){1}}     
\put(3,6){\line(1,0){1}}     
\put(5,6){\line(1,0){2}}     
\put(10,6){\line(1,0){4}}     
\put(2,2){\line(1,0){2}}     
\put(2.5,1.5){\line(1,0){1}}     
\put(14,1){\line(0,1){2}}     
\put(3,5){\line(0,1){1}}     
\put(6,1){\line(0,1){2.0}}     
\put(6,4.5){\line(0,1){1.5}}     
\put(14,1){\line(0,1){2}}     
\put(8,5.5){\line(0,1){0.5}}     
\put(9,5.5){\line(0,1){0.5}}     
\put(11,4){\line(0,1){2}}     
\put(14,5){\line(0,1){1}}     
\put(3,2){\line(0,1){1}}     
\put(11,1){\line(0,1){2}}     
\put(3,1){\line(0,1){0.5}}     
\put(4,6){\line(1,1){1}}     
\put(13.5,3){\line(0,1){2}}     
\put(14.5,3){\line(0,1){2}}     
\put(5.5,3){\line(1,3){0.5}}     
\put(6.5,3){\line(-1,3){0.5}}     
\put(7.5,6){\oval(1,1)[t]}       
\put(8.5,6){\oval(1,1)[t]}       
\put(9.5,6){\oval(1,1)[t]}      
\put(3,4){\oval(2,2)}                      
\put(4,6){\line(1,2){1}}     
\put(4,7.5){\makebox(0,0){$S$}}
\put(8.5,7.5){\makebox(0,0){$L$}}
\put(13,5){\makebox(0,0){$R$}}
\put(0.7,4){\makebox(0,0){$E$}}
\put(12,5){\makebox(0,0){$C$}}
\put(0.5,1.5){\makebox(0,0){$\Delta v$}}
\put(6.5,7.7){\makebox(0,0){$i$}}
\put(15.5,4.5){\makebox(0,0){$v$}}
\put(6,7){\vector(1,0){1}}     
\put(1.5,3.5){\vector(0,1){1}}     
\put(15,4){\vector(0,1){1}}     
\put(1.5,1){\vector(0,1){1}}     
}
\end{picture}
}
\caption{The three feedback schemes for time-delay autosyncronization of the \textit{buck} converter}
\label{fbuck}
\end{figure}

Figure \ref{fbuck} shows the basic scheme of the PWM controlled \textit{buck} converter together with the three
different schemes that we have studied to implement ETDAS. $v(t)$ is compared
with a periodic ramp
$$
r(t)=V_{\mbox{ref}}+\frac{V_L}{\sigma}+\frac{V_U-V_L}{\sigma T}t,
$$
with $t$ evaluated $\mbox{mod } T$, and the switch $S$ is opened if $v(t)>r(t)$
and closed if $v(t)<r(t)$.
If we assume that the inductor current $i(t)$ is always positive, the system has only two topologies 
and is described by 
$$
\frac{\diff v}{\diff t} = -\frac{1}{RC} v + \frac{1}{C} i,\ \ \ \ 
\frac{\diff i}{\diff t} = -\frac{1}{L} v +\frac{E}{L} u(t) 
$$
where $u(t)=1-\theta(v(t)-r(t))$. The input voltage $E$ acts as a bifurcation parameter. Notice that the system is nonlinear only
because of the changing topology.
We refer to \cite{DH}, \cite{diB} and \cite{FO} for a detailed description  of the bifurcation diagramm of this system and the chaotic regime.
In all the numerical simulations we will use the values $R=22\ \Omega$, $C=47\ \mu\mbox{F}$, $L=20\ \mbox{mH}$, $\sigma=8.4$,
$T=400\ \mu\mbox{s}$, $V_L=3.8 \ \mbox{V}$, $V_U=8.2\ \mbox{V}$, $V_{\mbox{ref}}=11.3\ \mbox{V}$ and $E$ varying in the range $[20,35]\ 
\mbox{V}$.

The control signal $\Delta v(t)$ is given by
$$
\Delta v(t) = \eta \left( v(t)-(1-r)\sum_{k=1}^{+\infty} r^{k-1} v(t-k\tau)\right),
$$
where $\eta$ is a (dimensionless) feedback gain and $r\in[0,1)$ determines the relative weight of the increasingly delayed contributions.
The case $r=0$ corresponds to TDAS. $\tau > 0$ is the period of the target UPO which, in our case, is always a multiple of the ramp period
$T$. Notice that $\Delta v(t)\equiv 0$ if $v(t)$ is periodic with period $\tau$.

We will be primarily concerned with the stabilization of UPOs which cross the ramp exactly once every period, and such that at the beginning
of every period one has $v(t)>r(t)$. More general situations can be easily treated as well, as will
become clear in Section \ref{buck_dc}.

\section{ETDAS for variable structure systems}
\label{d_of_c}
The PWM controlled \textit{buck} converter
described in Section \ref{buck}, as well as other PWM controlled DC-DC converters, can be considered as particular cases of
dynamical systems with equations of the form 
\begin{equation}
\dot x(t) = A(t,[x]) x(t) + b(t,[x]),
\label{eq1}
\end{equation} 
where $x, b \in\R^n$, $A\in\mbox{M}(\R^n)$. We use the notations  $A(t,[x])$ and $b(t,[x])$ to indicate
that both $A$ and $b$ are local functionals of $x$. For the PWM controlled converters, $A$ and
$b$ are piecewise constant, tipically changing their values when a linear function of $x$ crosses a given
periodic function of $t$.

Consider now a $T-$periodic orbit $x^*(t)$ of this system and a nearby orbit $x(t)$. We want to study the
evolution of $y(t)=x(t)-x^*(t)$. We will be mainly interested in unstable orbits $x^*(t)$, and our goal will 
be to modify the right-hand side of (\ref{eq1}) so as to render $x^*(t)$ stable, 
that is $\lim_{t\to+\infty}y(t)=0$ for $x(t)$ initially close enough to $x^*(t)$. We will do so by means of
extended time-delay autosynchronization and consider the following equation:
\begin{equation}
\dot x(t) = A(t,[x]) x(t) + b(t,[x])+\eta M(t,[x]) (x(t)-(1-r)\sum_{k=1}^{+\infty}r^{k-1}x(t-k\tau)),
\label{eq2}
\end{equation} 
where $M\in\mbox{M}(\R^n)$ is a matrix indicating how the delayed signal is feeded back to the system and
$\eta$ is the strength of the feedback gain.

We will study the evolution of $y(t)$ to first order in $y$ under (\ref{eq2}). One has
\begin{eqnarray*}
\dot y(t) 
&=&  A(t,[x])x(t)-A(t,[x^*])x^*(t) + b(t,[x]) - b(t,[x^*])\\
&+&\eta M(t,[x^*])  (y(t)-(1-r)\sum_{k=1}^{+\infty}r^{k-1}y(t-k\tau)) + O(y^2)
\end{eqnarray*}
where the $\tau-$periodicity of $x^*$ has been used and we have thrown away terms of higher order in $y$. Expanding
the functionals around $x^*$ gives
\begin{eqnarray}
\dot y(t) &=& A(t,[x^*]) y(t) + \left[\int_0^\tau 
\left.\frac{\delta A(t,[x])}{\delta x_k(t')}\right|_{x=x^*} y_k(t') \mbox{ d}t' \right] x^*(t) 
+\int_0^\tau 
\left.\frac{\delta b(t,[x])}{\delta x_k(t')}\right|_{x=x^*} y_k(t') \mbox{ d}t' \nonumber \\
&+& \eta M(t,[x^*])  (y(t)-(1-r)\sum_{k=1}^{+\infty}r^{k-1}y(t-k\tau)) + O(y^2).
\label{eq3p}
\end{eqnarray}
Since the functionals are local, the functional derivatives will
yield delta functions and, forgetting about the higher order terms, we get a variable coefficient, 
linear time-delayed differential equation for $y$ which, with suitable definitions,
can be written in the form
\begin{equation}
\dot y(t) = A_0(t) y(t) + \sum_{k=1}^{+\infty} A_k(t) y(t-k\tau),
\label{eq3}
\end{equation}
where the time dependence of the coefficients is known since $x^*(t)$ is known. Notice also that these coefficients are
periodic functions of time, since their explicit dependence of time is periodic and $x^*$ is also periodic.
From (\ref{eq3p}) and (\ref{eq3}) one can easily read
\begin{equation}
A_k(t)=-(1-r)\eta M(t,[x^*]) r^{k-1},\ \ \ k=1,2,\ldots.
\label{eq4}
\end{equation}

We are not interested in the general solution of the time-delayed equation (\ref{eq3}), but rather we would like to know
if its zero solution is assimptotically stable. To this end, we look for solutions of the form
$$
y(t)=p_\lambda(t) e^{\lambda t/\tau},
$$
with $\lambda\in\C$ and $p_\lambda(t+\tau)=p_\lambda(t)$. This yields for $p_\lambda$ an ordinary differential equation
$$
\dot p_\lambda(t) = (A_0(t)-\frac{\lambda}{\tau}\I)p_\lambda(t) + \sum_{k=1}^{+\infty} e^{-k\lambda} A_k(t) p_\lambda(t),
$$
whose solution for a given initial condition can be expressed in terms of an evolution operator $U_\lambda(t)$ defined
by
$$
p_\lambda(t)=e^{-\lambda t/\tau} U_\lambda(t) p_\lambda(0)
$$
and satisfying the equation
\begin{equation}
\dot U_\lambda(t) = \left(A_0(t)+\sum_{k=1}^{+\infty} e^{-k\lambda} A_k(t)\right) U_\lambda(t)
\label{eq5}
\end{equation}
with initial condition $U_\lambda(0)=\I$.
The general solution to (\ref{eq5}) can be formally expressed as
\begin{equation}
U_\lambda(t) = {\cal T} \exp\left(\int_0^t (A_0(t')+\sum_{k=1}^{+\infty} e^{-k\lambda} A_k(t'))\ \mbox{d}t'\right),
\label{eq6}
\end{equation}
where ${\cal T}$ stands for time ordered product (this formal solution is also known as Peano-Baker series
in the mathematical literature \cite{Rugh}; it boils down to the standard exponential matrix if the coefficients of the differential
equation are constant). In any case, $U_\lambda(t)$ retains the fundamental properties 
\begin{equation}
U_\lambda(t_1+t_2)=U_\lambda(t_1)U_\lambda(t_2),\ \ \ U_\lambda(-t)=U_\lambda^{-1}(t).
\label{eq7}
\end{equation}
Using (\ref{eq7}) the condition $p_\lambda(t+\tau)=p_\lambda(t)$ is easily seen to be equivalent to
$$
\left(e^{-\lambda}U_\lambda(\tau)-\I\right)p_\lambda(0) = 0
$$
which implies
\begin{equation}
\det \left(e^{-\lambda}U_\lambda(\tau)-\I\right)=0
\label{eq8}
\end{equation}
This equation determines the values of $\lambda$ such that $p_\lambda(t)$ is a solution of our equation. As we want 
$y(t)=e^{\lambda t/\tau}p_\lambda(t)$ go asymptotically to zero, we must demand that $\Re\lambda<0$ for all 
the solutions of  (\ref{eq8}).
Defining the Floquet multiplier $\mu=e^\lambda$ and $U(\tau;\mu^{-1})=U_\lambda(\tau)$, equation (\ref{eq8}) becomes
\begin{equation}
g(\mu^{-1})\equiv\det(\mu^{-1} U(\tau;\mu^{-1})-\I)=0
\label{eq9}
\end{equation}
with
\begin{equation}
U(\tau;\mu^{-1}) = {\cal T} \exp\left(\int_0^\tau (A_0(t)+\sum_{k=1}^{+\infty} \mu^{-k} A_k(t))\ \mbox{d}t \right).
\label{eq10}
\end{equation}
Summing up, stability of $x^*$ is equivalent to the requirement that all the zeros of $g(\mu^{-1})$ lie outside the unit
circle ($ ||\mu^{-1}|| > 1 \Leftrightarrow \Re\lambda <0$). Using (\ref{eq4}) the series in (\ref{eq10}) can be summed
to yield
\begin{equation}
U(\tau;\mu^{-1})={\cal T} \exp\left(\int_0^\tau (A_0(t)-(1-r)\eta M(t,[x^*])\frac{\mu^{-1}}{1-r\mu^{-1}})\ \mbox{d}t\right).
\label{eq10b}
\end{equation}
For $r<1$, $U(\tau;\mu^{-1})$ as a function of $\mu^{-1}$, and hence $g(\mu^{-1})$, has no poles inside the unit circle.
By a well-known theorem of complex analysis, the number of zeros of $g(\mu^{-1})$ inside the unit circle equals the
index with respect to the origin of the curve traced by $g(\mu^{-1})$ when $\mu^{-1}$ runs over the unit circle. Therefore, 
the solution will be stable if and only if this index is zero. This way of computing the number of zeros inside the
unit circle is numerically preferred in front of the obvious option of actually computing the zeros, which in fact
are infinite (this is due to the time-delay character of the original equation). 

The equations derived so far are completely general. In practice, however, analytical computation
 of $U(\tau;\mu^{-1})$ using (\ref{eq10b}) is not possible except for very special
cases, so one must fall back to the defining differential equation (\ref{eq5}) and numerically integrate it between
$0$ and $\tau$. For the \textit{buck} converter, the functional derivatives appearing in (\ref{eq3p}) are specially simple (in fact,
$A$ is constant), and the differential equation (\ref{eq5}) can be analytically integrated. We will do so in the next
section, for several choices of $M$.

\section{Analytical computation of $g(\mu^{-1})$ for the \textit{buck} converter}
\label{buck_dc}
For the \textit{buck} converter, one has 
$$
A(t,[x])=\left(\begin{array}{cc}
             -\frac{1}{RC} & \frac{1}{C} \\
             -\frac{1}{L}  & 0 
             \end{array}
            \right) \equiv A
\ \
\mbox{and} 
\ \
b(t,[x])=\left(\begin{array}{c}
           0 \\ \frac{E}{L}
           \end{array}
         \right) (1-\theta(v(t)-r(t)))
$$
so that all the functional derivatives of $A(t,[x])$ are zero and the only nonzero functional derivative of $b(t,[x])$ is
$$
\frac{\delta b_2(t,[x])}{\delta v(\tau)} = -\frac{E}{L}\delta(v(t)-r(t))\delta(t-\tau).
$$
Then, equation (\ref{eq3p}) becomes, with $\tau=T$,
\begin{equation}
\dot y(t) = A y(t)  
-\frac{E}{L} \left(\begin{array}{cc} 0 & 0 \\ 1 & 0 \end{array} \right) y(t)  \delta(v^*(t)-r(t))
+ \eta M(t,[x^*])  \left(y(t)-(1-r)\sum_{k=1}^{+\infty}r^{k-1}y(t-kT)\right)
\label{eq11}
\end{equation}

It is easy to see that the three feedback schemes described in Section \ref{buck} correspond to the following choices
for the matrix $M$:
\begin{enumerate}
\item 
$$
M(t,[x]) = \left(\begin{array}{cc}
             \frac{1}{RC} & 0 \\
             0  & 0 
             \end{array}
            \right)\equiv M_1
$$
\item
$$
M(t,[x]) = \left(\begin{array}{cc}
             -\frac{1}{RC} & 0 \\
             -\frac{1}{L}  & 0 
             \end{array}
            \right)  \equiv M_2         
$$
\item
$$
M(t,[x]) = (1-\theta(v(t)-r(t)) 
           \left(\begin{array}{cc}
             0 & 0 \\
             \frac{1}{L}  & 0 
             \end{array}
            \right) \equiv  (1-\theta(v(t)-r(t))\ M_3
$$
\end{enumerate}
Integration of the resulting differential equation is very similar for cases 1 and 2, while 3 is slightly more
involved. We will present the explicit computation of $g(\mu^{-1})$ for cases 1 and 3 and deduce the result for
case 2 from the one obtained in case 1.

In the first feedback scheme one has
\begin{eqnarray*}
A_0(t) &=& A +\eta M_1-\frac{E}{L} \left(\begin{array}{cc} 0 & 0 \\ 1 & 0 \end{array} \right)\delta(v^*(t)-r(t))\\
A_k(t) &=& -\eta (1-r) r^{k-1} M_1,\ \ k=1,2,\ldots
\end{eqnarray*}
and the differential equation for $U(t)$ is, adding up the geometrical series,
\begin{equation}
\dot U(t) = \left(\begin{array}{cc}
   -\frac{1}{RC}(1-\eta+\eta \mu^{-1}\frac{1-r}{1-r\mu^{-1}}) & \frac{1}{C} \\
   -\frac{1}{L}-\frac{E}{L} \delta(v^*(t)-r(t)) & 0 
   \end{array}
   \right) U(t)
 \label{eq13}  
 \end{equation}
   
If we write
$$
U(t)=\left( \begin{array}{cc}
         u_1(t) & u_2(t) \\
         u_3(t) & u_4(t) 
         \end{array}
     \right)
$$
we get two uncoupled systems of dimension two
\begin{eqnarray*}
\dot u_1(t) &=& -a_1 u_1(t) + \frac{1}{C} u_3(t)\\
\dot u_2(t) &=& -a_1 u_2(t) + \frac{1}{C} u_4(t)\\           
\dot u_3(t) &=& -\frac{1}{L} u_1(t)-\frac{E}{L}\delta(v^*(t)-r(t))u_1(t)\\             
\dot u_4(t) &=& -\frac{1}{L} u_2(t)-\frac{E}{L}\delta(v^*(t)-r(t))u_2(t)
\end{eqnarray*}
where 
$$
a_1 = \frac{1}{RC} \left(1-\eta+\eta \mu^{-1}\frac{1-r}{1-r\mu^{-1}}\right),
$$
with the initial conditions $u_1(0)=1$, $u_2(0)=0$, $u_3(0)=0$, $u_4(0)=1$. Therefore, we only need to solve
twice the single system
\begin{eqnarray}
\dot x(t) &=& -a_1 x(t) + \frac{1}{C} y(t)\nonumber \\
\dot y(t) &=& -\frac{1}{L} x(t)-\frac{E}{L}\delta(v^*(t)-r(t))x(t)
\label{syst}
\end{eqnarray}
for $t\in [0,T]$ with the two sets of initial conditions $(1,0)$ and $(0,1)$.
If we assume that for each cycle of the auxiliary signal $r(t)$ there is one and only one
$t=t_c\in(0,T)$ such that the system switches topology, then the delta function appearing in
the above equation is
$$
\delta(v^*(t)-r(t))=\beta \delta(t-t_c),
$$
where
$$
\beta=\frac{1}{|\dot v^*(t_c)-\dot r(t_c)|}
$$
is the inverse of the absolute value of the slope with which $v^*(t)$ crosses the ramp. This is well
defined because $v(t)$, in contrast to $i(t)$, which is not differentiable at $t=t_c$, 
is everywhere differentiable. We get thus the time-varying linear system
\begin{eqnarray}
\dot x(t) &=& -a_1 x(t) + \frac{1}{C} y(t) \label{eqs1a}\\
\dot y(t) &=& -\frac{1}{L} x(t)-\frac{\beta E}{L}\delta(t-t_c)x(t)
\label{eqs1b}
\end{eqnarray}
Although this is a variable coefficient system, it can be solved by Laplace transform since the delta
function makes it trivial to compute the transform of the last term of (\ref{eqs1b}). We skip the
details and write down the general solution for $t\in[0,T]$:
\begin{eqnarray*}
x(t)&=& \left(-x(0)\frac{a_1-\gamma_1}{2\gamma_1}+y(0)\frac{1}{\gamma_1 C}\right) e^{-\frac{1}{2}(a_1-\gamma_1)t} 
    + \left(x(0)\frac{a_1+\gamma_1}{2\gamma_1}-y(0)\frac{1}{\gamma_1 C}\right) e^{-\frac{1}{2}(a_1+\gamma_1)t} \\
    &-&\frac{E\beta}{LC\gamma_1}x(t_c)\theta(t-t_c)\left(e^{-\frac{1}{2}(a_1-\gamma_1)(t-t_c)}-e^{-\frac{1}{2}(a_1+\gamma_1)(t-t_c)}\right)\\
y(t)&=& \left(-x(0)\frac{1}{L\gamma_1}+y(0)\frac{a_1+\gamma_1}{2\gamma_1}\right) e^{-\frac{1}{2}(a_1-\gamma_1)t} 
    + \left(x(0)\frac{1}{L\gamma_1}-y(0)\frac{a_1-\gamma_1}{2\gamma_1}\right) e^{-\frac{1}{2}(a_1+\gamma_1)t} \\
    &-&\frac{E\beta}{2L\gamma_1}x(t_c)\theta(t-t_c)
    \left((a_1+\gamma_1)e^{-\frac{1}{2}(a_1-\gamma_1)(t-t_c)}-(a_1-\gamma_1)e^{-\frac{1}{2}(a_1+\gamma_1)(t-t_c)}\right)
\end{eqnarray*} 
where
$$
\gamma_1=\sqrt{a_1^2-\frac{4}{LC}}
$$
and $x(t_c)$ is defined evaluating the expression for $x(t)$ at $t=t_c$. Notice that $x(t)$ is continuous at $t=t_c$, while
$y(t)$ has a jump of value $-\beta E/L x(t_c)$, as follows from (\ref{eqs1b}). 

ow putting $x(0)=1$, $y(0)=0$ we obtain, at $t=T$, $u_1(T)$ and $u_3(T)$, while $x(0)=0$ and $y(0)=1$ yield us
$u_2(T)$ and $u_4(T)$. Then we can compute
\begin{eqnarray}
g(\mu^{-1})&=&\det\left|\begin{array}{cc}
     \mu^{-1} u_1(T)-1 & \mu^{-1} u_2(T) \\
     \mu^{-1} u_3(T) & \mu^{-1}u_4(T)-1
     \end{array}
     \right| \nonumber \\  
 &=& \mu^{-2}\left(u_1(T)u_4(T)-u_2(T)u_3(T)\right)-\mu^{-1}\left(u_1(T)+u_4(T)\right)+1
\label{pf}
\end{eqnarray}
which, after a little algebra, yields
\begin{equation}
g_1(\mu^{-1})=
\mu^{-2}e^{-a_1T}-2\mu^{-1}e^{-\frac{1}{2}a_1T}\left(\cosh(\frac{1}{2}\gamma_1 T)-\frac{E\beta}{\gamma_1 LC}\sinh(\frac{1}{2}\gamma_1 T)
\right) +1
\label{eqs2}
\end{equation}
The term that multiplies $\mu^{-2}$ could have been easily computed without solving the system, since it equals the Wronskian, $W$,
 and satisfies $\dot W = -a_1 W$, with $W(0)=1$.

Notice that, although $\gamma_1$ introduces a line cut where its argument vanishes, if we analytically extend the above expression
by its series expansion, only even powers of $\gamma_1$ do appear and thus $g_1(\mu^{-1})$ only has pole-type singularities (outside
the unit circle for $r<1$). 

In case 2, the equation obeyed by $U(t)$ is
$$
\dot U(t) = \left(\begin{array}{cc}
   -\frac{1}{RC}(1+\eta-\eta \mu^{-1}\frac{1-r}{1-r\mu^{-1}}) & \frac{1}{C} \\
   -\frac{1}{L}(1+\eta-\eta \mu^{-1}\frac{1-r}{1-r\mu^{-1}})-\frac{E}{L} \delta(v^*(t)-r(t)) & 0 
   \end{array}
   \right) U(t)  
$$             
Comparing with case 1, the differences amount to changing the sign of $\eta$ and then replacing $1/L$ by
$1/L(1+\eta-\eta \mu^{-1}\frac{1-r}{1-r\mu^{-1}})$ without changing the combination $E/L$. Thus, the index function in this case is
\begin{equation}
g_2(\mu^{-1})=
\mu^{-2}e^{-a_2T}-2\mu^{-1}e^{-\frac{1}{2}a_2T}\left(\cosh(\frac{1}{2}\gamma_2 T)-\frac{E\beta}{\gamma_2 LC}\sinh(\frac{1}{2}\gamma_2 T)
\right) +1
\label{eqs4}
\end{equation}
with
$$
a_2 = \frac{1}{RC} \left(1+\eta-\eta \mu^{-1}\frac{1-r}{1-r\mu^{-1}}\right)
$$ 
and
$$
\gamma_2=\sqrt{a_2^2-\frac{4}{LC}\left(1+\eta-\eta\mu^{-1}\frac{1-r}{1-r\mu^{-1}}\right)}.
$$

For the third scheme the evolution operator $U(t)$ obeys the equation
$$
\dot U(t) = \left(\begin{array}{cc}
   -\frac{1}{RC} & \frac{1}{C} \\
   & \\
   -\frac{1}{L}-\frac{E}{L} \delta(v^*(t)-r(t))+\frac{\eta}{L}(1-\theta(v^*(t)-r(t))\left(1-\mu^{-1}\frac{1-r}{1-r\mu^{-1}}\right) & 0 
   \end{array}
   \right) U(t)  
$$    

Following the same steps as in the previous cases, we arrive at the time-varying linear system
\begin{eqnarray*}
\dot x(t) &=& -\frac{1}{RC} x(t) + \frac{1}{C} y(t) \\
\dot y(t) &=& -\frac{1}{L} x(t)-\frac{\beta E}{L}\delta(t-t_c)x(t)+\frac{\eta}{L}\theta(t-t_c)(1-\mu^{-1}\frac{1-r}{1-r\mu^{-1}})x(t),
\end{eqnarray*}
which we must solve for the initial conditions $(1,0)$ and $(0,1)$. For $t_c < t\leq T$ the solution is
\begin{eqnarray*}
x(t)&=& \left(-x(t_c)\frac{a-\gamma_3}{2\gamma_3}+y(t_c)\frac{1}{\gamma_3 C}\right) e^{-\frac{1}{2}(a-\gamma_3)(t-t_c)} \\
    &+& \left(x(t_c)\frac{a+\gamma_3}{2\gamma_3}-y(t_c)\frac{1}{\gamma_3 C}\right) e^{-\frac{1}{2}(a+\gamma_3)(t-t_c)} \\   
y(t)&=& \left(-x(t_c)\frac{1}{L_3\gamma_3}+y(t_c)\frac{a+\gamma_3}{2\gamma_3}\right) e^{-\frac{1}{2}(a-\gamma_3)(t-t_c)} \\ 
    &+& \left(x(t_c)\frac{1}{L_3\gamma_3}-y(t_c)\frac{a-\gamma_3}{2\gamma_3}\right) e^{-\frac{1}{2}(a+\gamma_3)(t-t_c)} \\    
\end{eqnarray*} 
where
\begin{eqnarray*}
x(t_c)&=& \left(-x(0)\frac{a-\gamma}{2\gamma}+y(0)\frac{1}{\gamma C}\right) e^{-\frac{1}{2}(a-\gamma)t_c} 
    + \left(x(0)\frac{a+\gamma}{2\gamma}-y(0)\frac{1}{\gamma C}\right) e^{-\frac{1}{2}(a+\gamma)t_c} \\
y(t_c)&=& \left(-x(0)\frac{1}{L\gamma}+y(0)\frac{a+\gamma}{2\gamma}\right) e^{-\frac{1}{2}(a-\gamma)t_c} 
    + \left(x(0)\frac{1}{L\gamma}-y(0)\frac{a-\gamma}{2\gamma}\right) e^{-\frac{1}{2}(a+\gamma)t_c}
    -\frac{E\beta}{L}x(t_c)
\end{eqnarray*} 
(here $y(t_c)$ really means $y(t_c^+)$) and
$$
a = \frac{1}{RC}, \ \
\gamma = \sqrt{a^2-\frac{4}{LC}}, \ \
\frac{1}{L_3} = \frac{1}{L}\left(1-\eta\left(1-\mu^{-1}\frac{1-r}{1-r\mu^{-1}}\right)\right), \ \
\gamma_3 = \sqrt{a^2-\frac{4}{L_3C}}.
$$

After some calculations one can get
\begin{equation}
g_3(\mu^{-1}) = \mu^{-2} e^{-aT} -2\mu^{-1} e^{-\frac{1}{2}aT} (A-\frac{E\beta}{LC} B) +1,                 
\label{eqs6}
\end{equation}
                                                
where

\begin{eqnarray*}
A &=& \cosh(\frac{1}{2}\gamma t_c)\cosh(\frac{1}{2}\gamma_3(T-t_c)) 
            +\frac{1}{\gamma\gamma_3}\left(a^2-2\frac{L+L_3}{CLL_3}\right)\sinh(\frac{1}{2}\gamma t_c)\sinh(\frac{1}{2}\gamma_3(T-t_c)) \\
B &=& \frac{1}{\gamma} \sinh(\frac{1}{2}\gamma t_c)\cosh(\frac{1}{2}\gamma_3(T-t_c))
            +\frac{1}{\gamma_3}\cosh(\frac{1}{2}\gamma t_c)\sinh(\frac{1}{2}\gamma_3(T-t_c))
\end{eqnarray*}                                                

The main new feature of this expression with respect to $g_1$ and $g_2$ is the explicit appearance of $t_c$. As a check, one can see that
$g_3(\mu^{-1})$ reduces to $g_1(\mu^{-1})$ if one sets $a_1=a$, $\gamma_3=\gamma_1=\gamma$, as it must be.

\section{Numerical analysis of  $g(\mu^{-1})$}
\label{buck_num}

\begin{figure}[ht]
\begin{center}
\subfigure[First scheme, $r=0.0$]{\epsfig{file=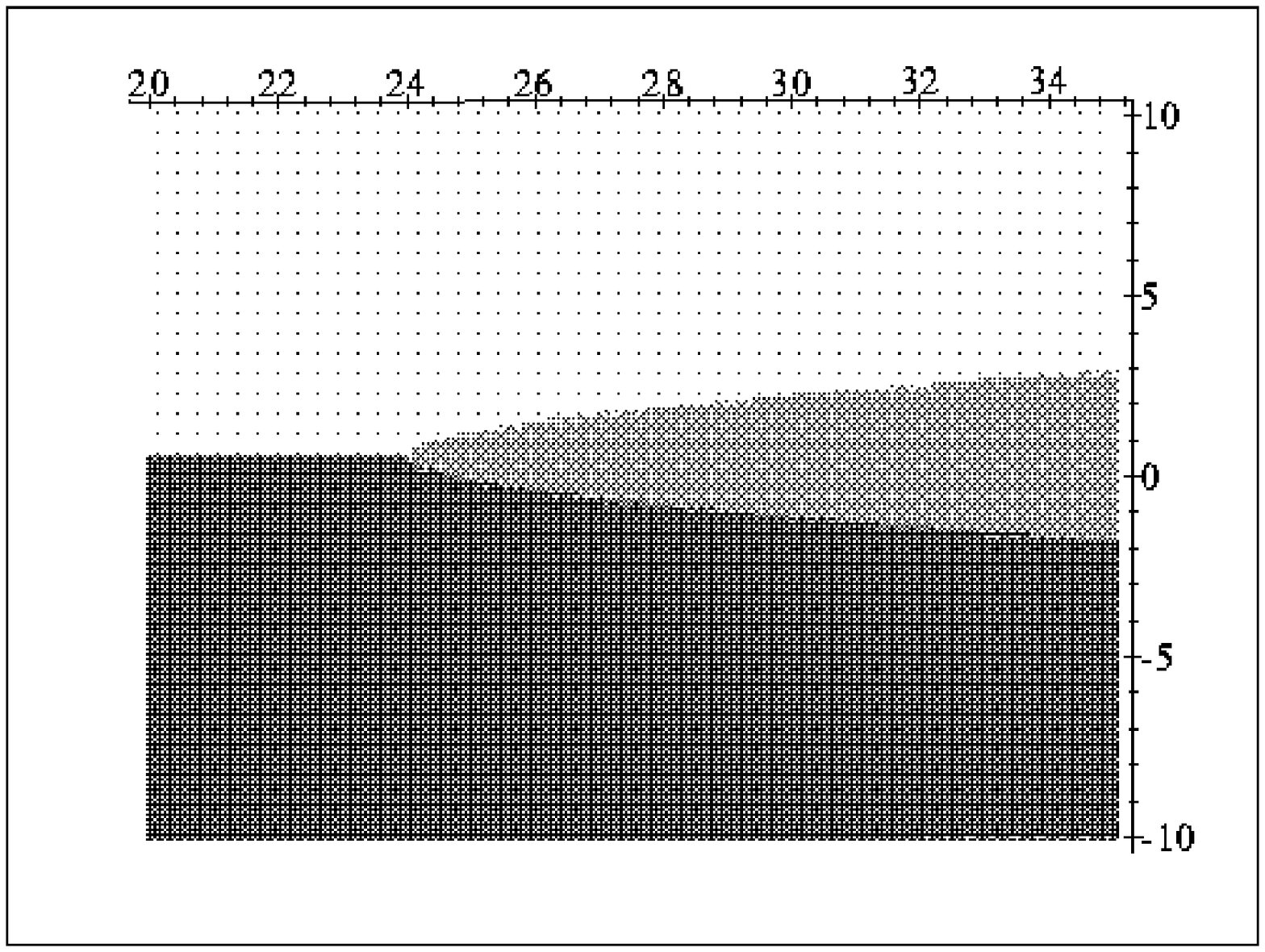,height=5cm,width=5cm,angle=-90}}
\subfigure[First scheme, $r=0.6$]{\epsfig{file=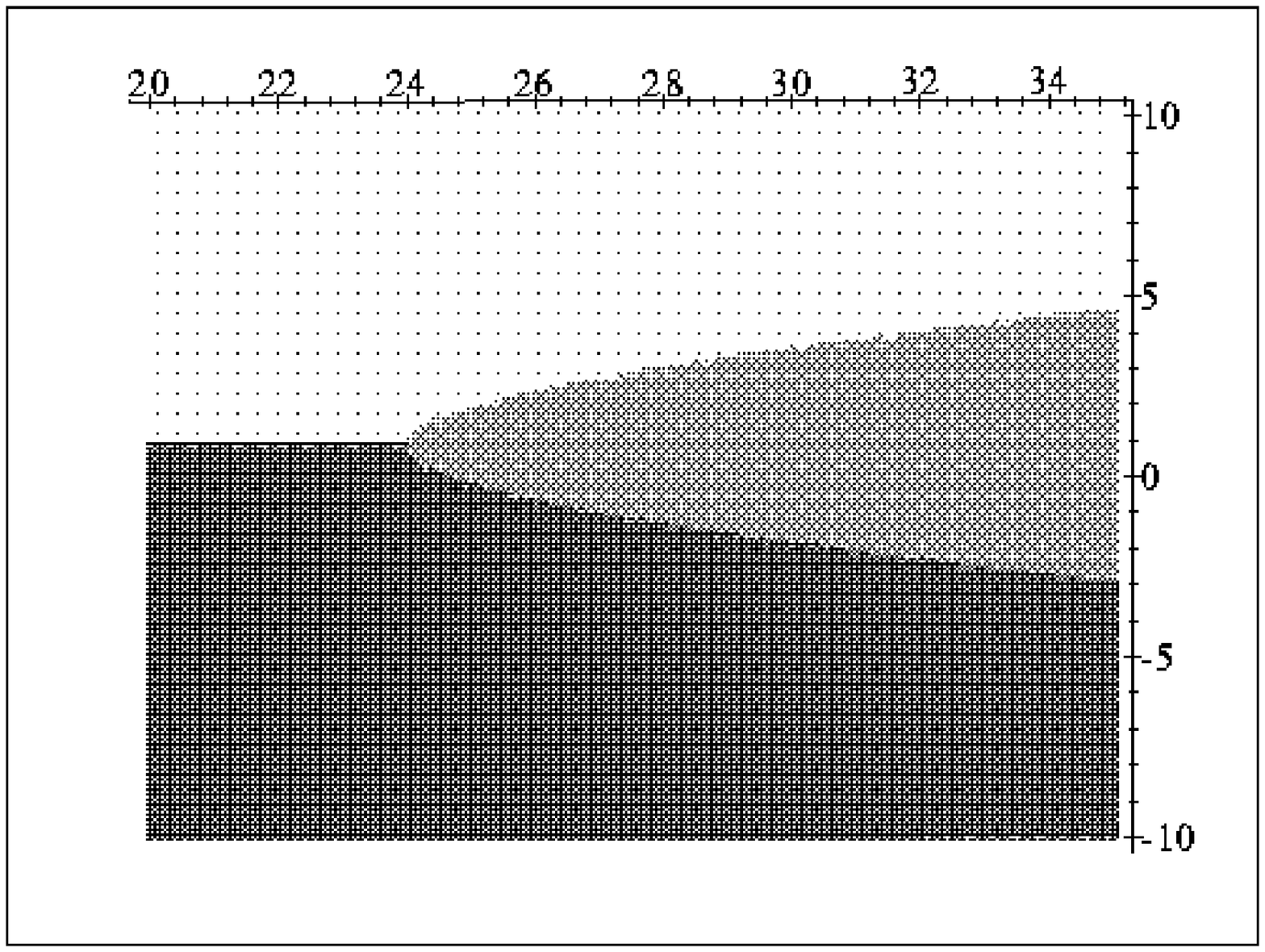,height=5cm,width=5cm,angle=-90}}
\subfigure[First scheme, $r=0.9$]{\epsfig{file=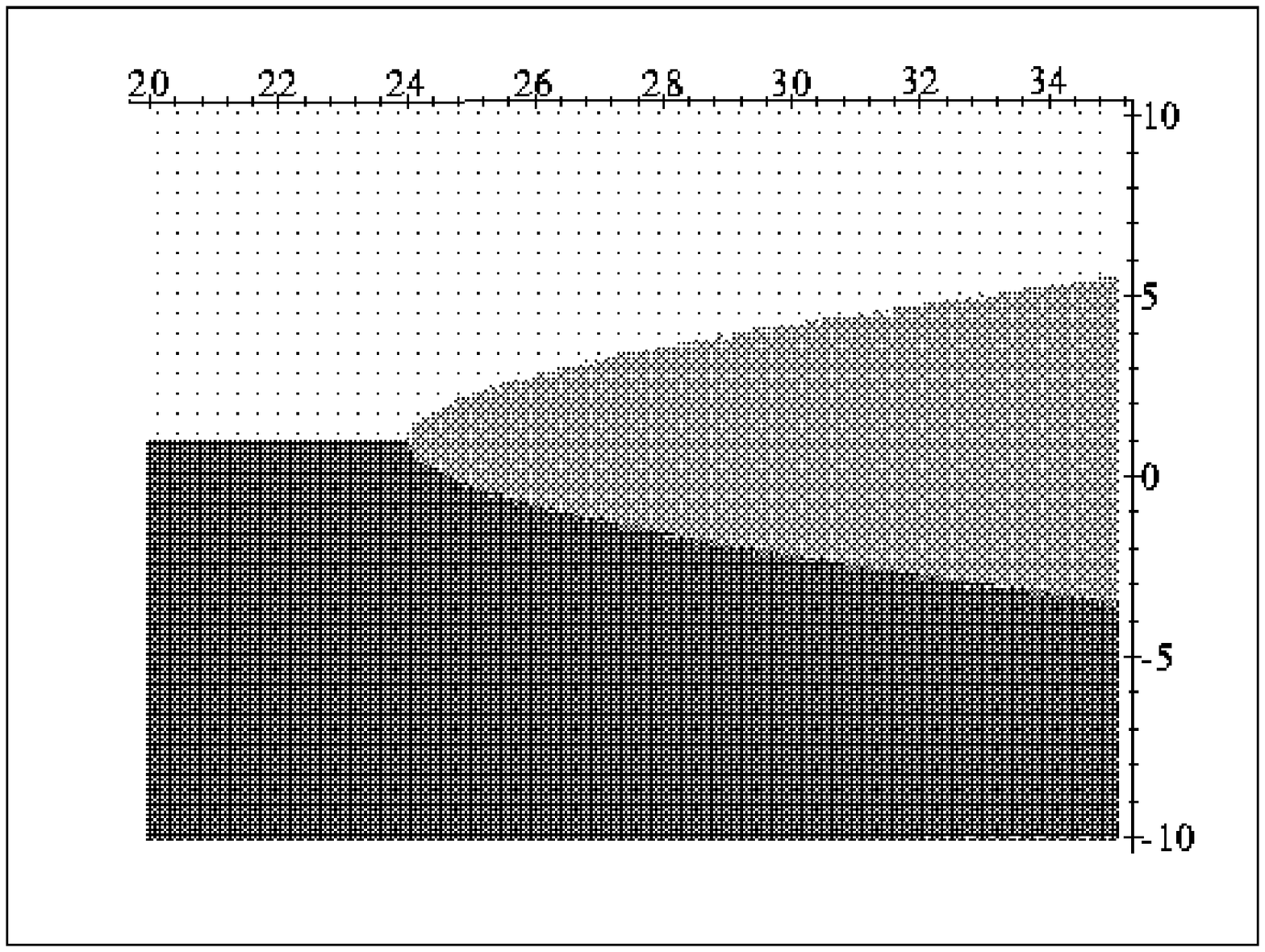,height=5cm,width=5cm,angle=-90}}

\subfigure[Second scheme, $r=0.0$]{\epsfig{file=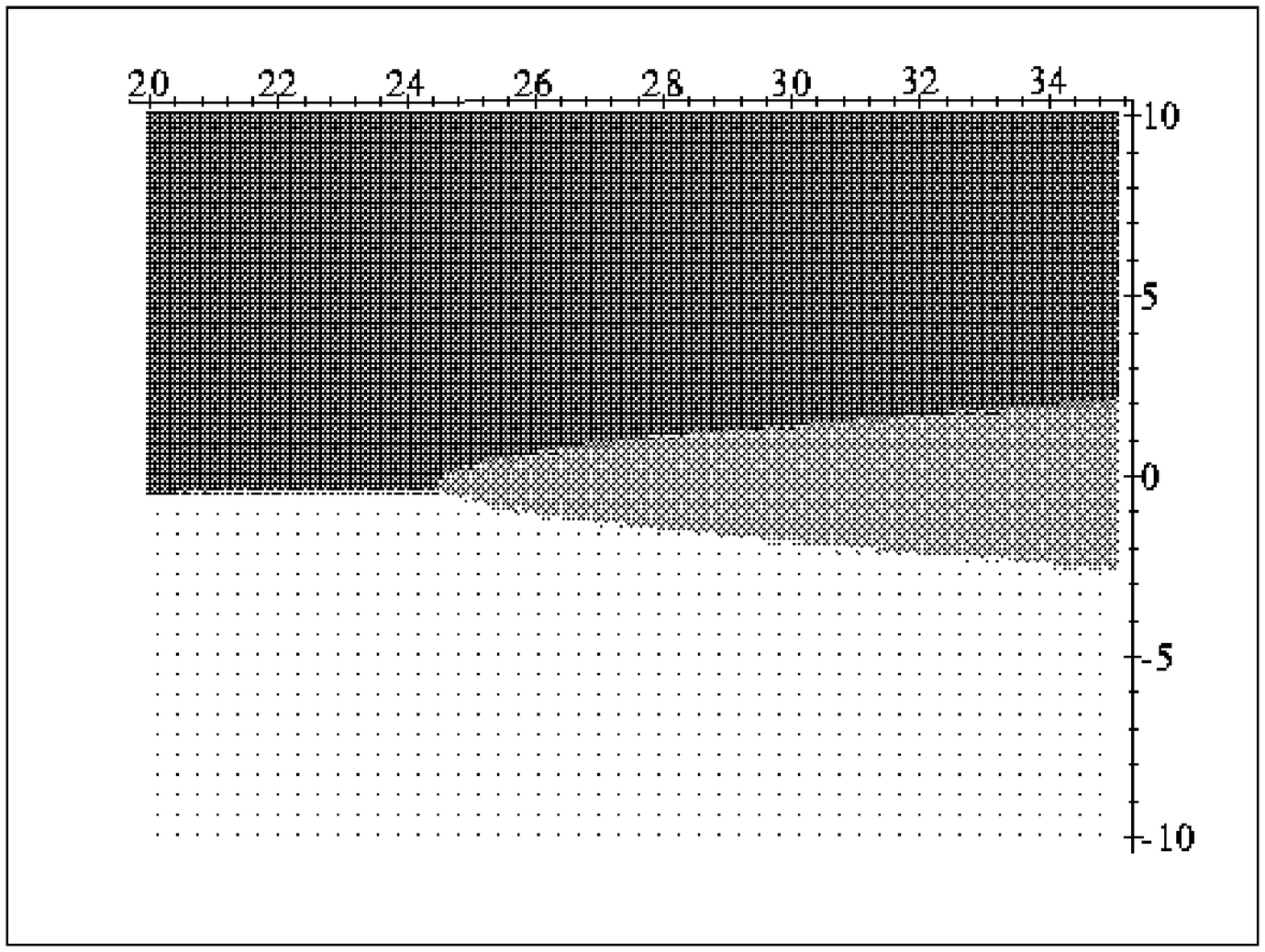,height=5cm,width=5cm,angle=-90}}
\subfigure[Second scheme, $r=0.6$]{\epsfig{file=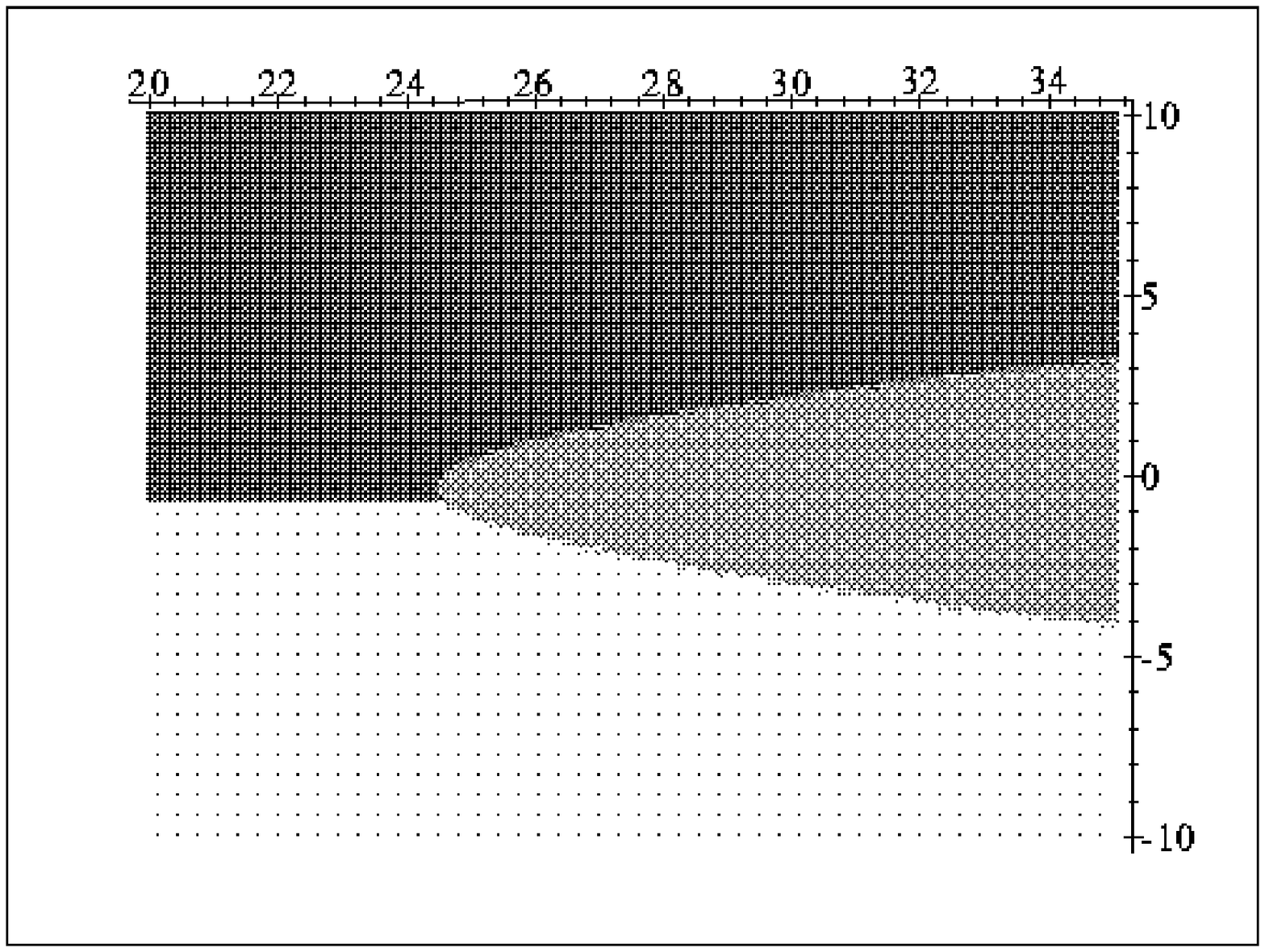,height=5cm,width=5cm,angle=-90}}
\subfigure[Second scheme, $r=0.9$]{\epsfig{file=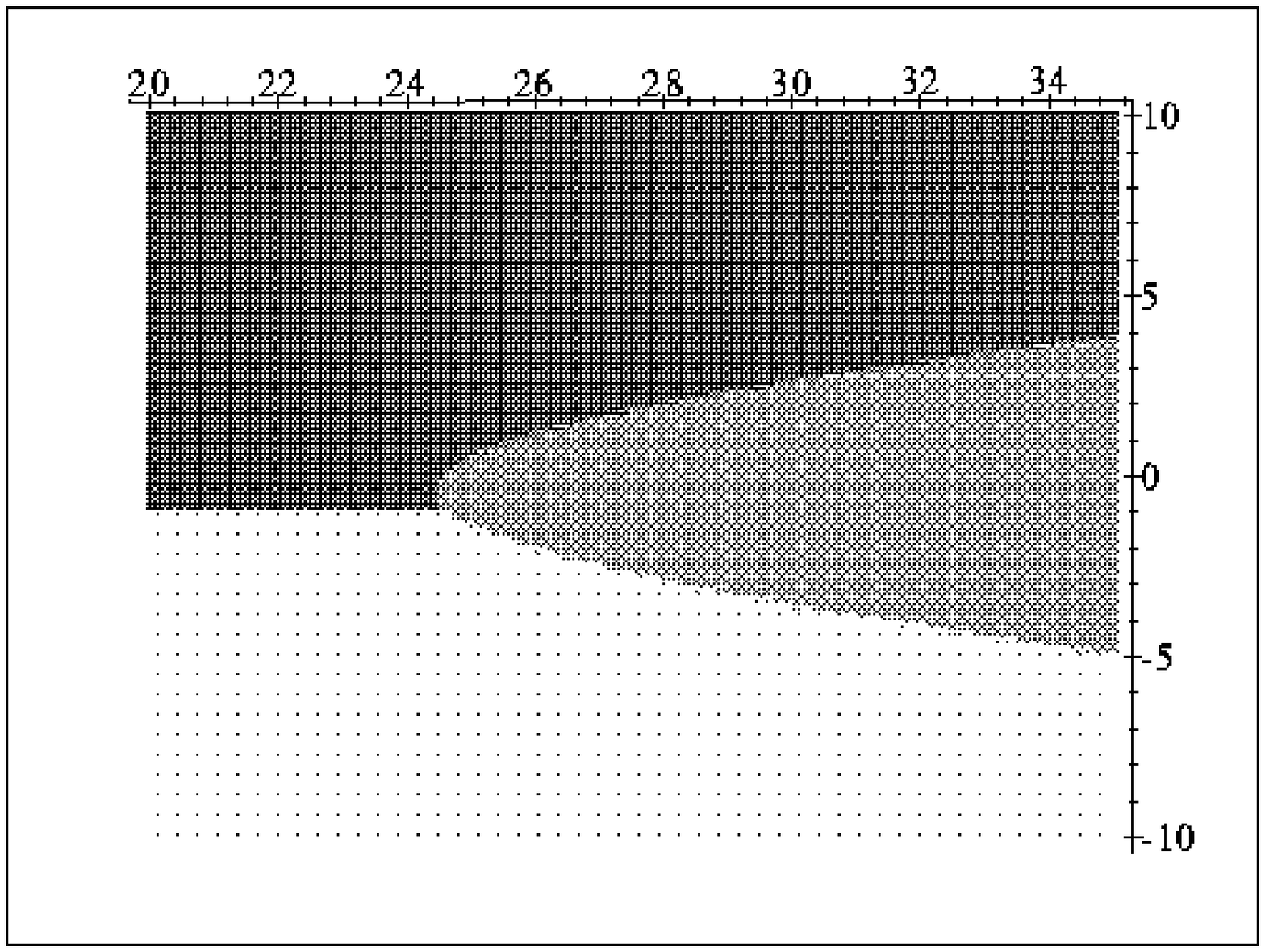,height=5cm,width=5cm,angle=-90}}

\caption{Domains of control for the two first feedback schemes and several values of $r$. Vertical axis: $\eta$ from $-10$ to $10$  . 
Horizontal axis: $E$ from $20$ V to $35$ V. 
Black $=0$ (stable), Grey $=1$ (unstable), White $=2$ (unstable).}
\label{dc12}
\end{center}
\end{figure}

\begin{figure}[ht]
\begin{center}
\subfigure[$r=0.0$]{\epsfig{file=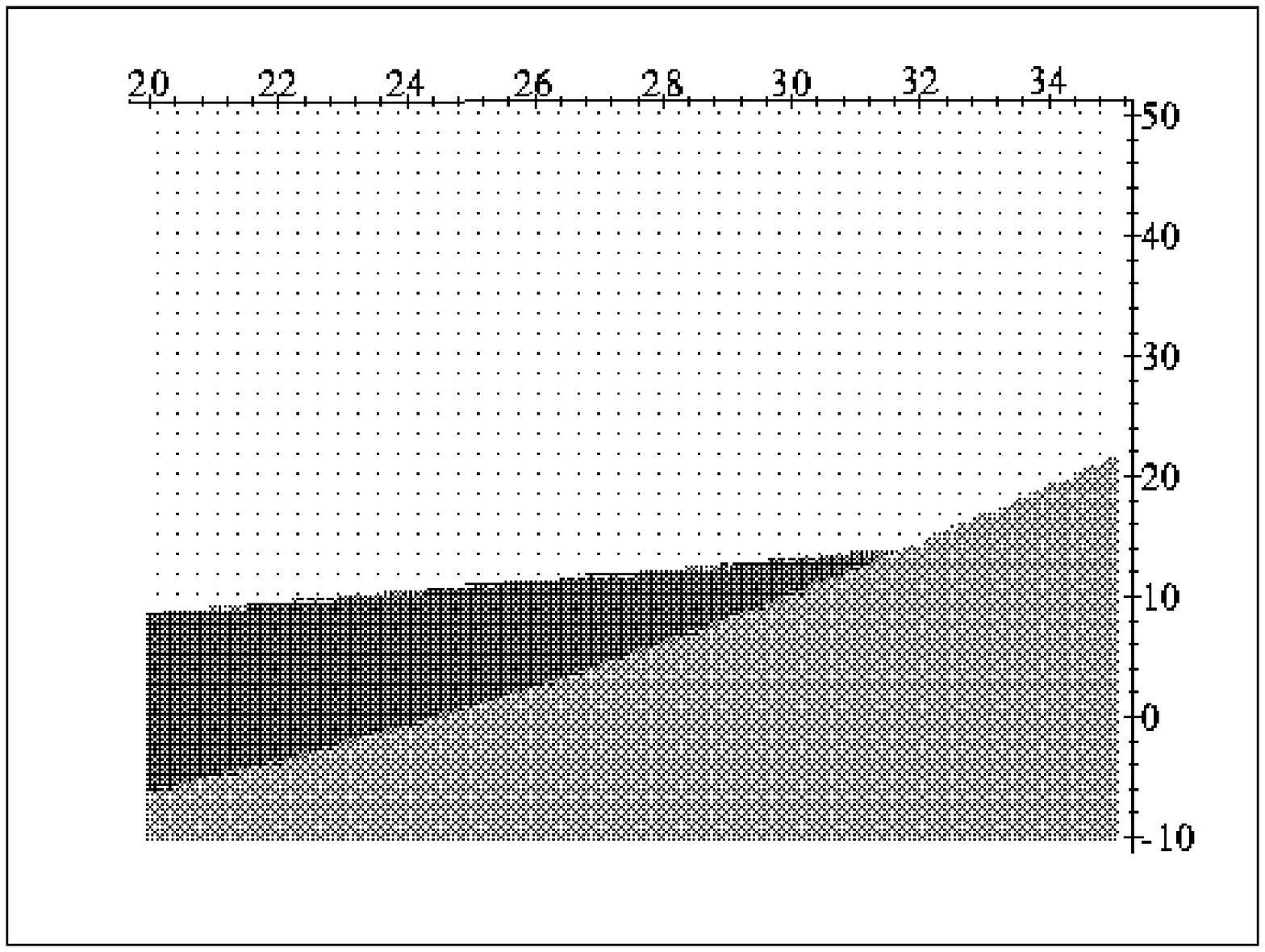,height=4.5cm,width=4.5cm,angle=-90}}
\subfigure[$r=0.6$]{\epsfig{file=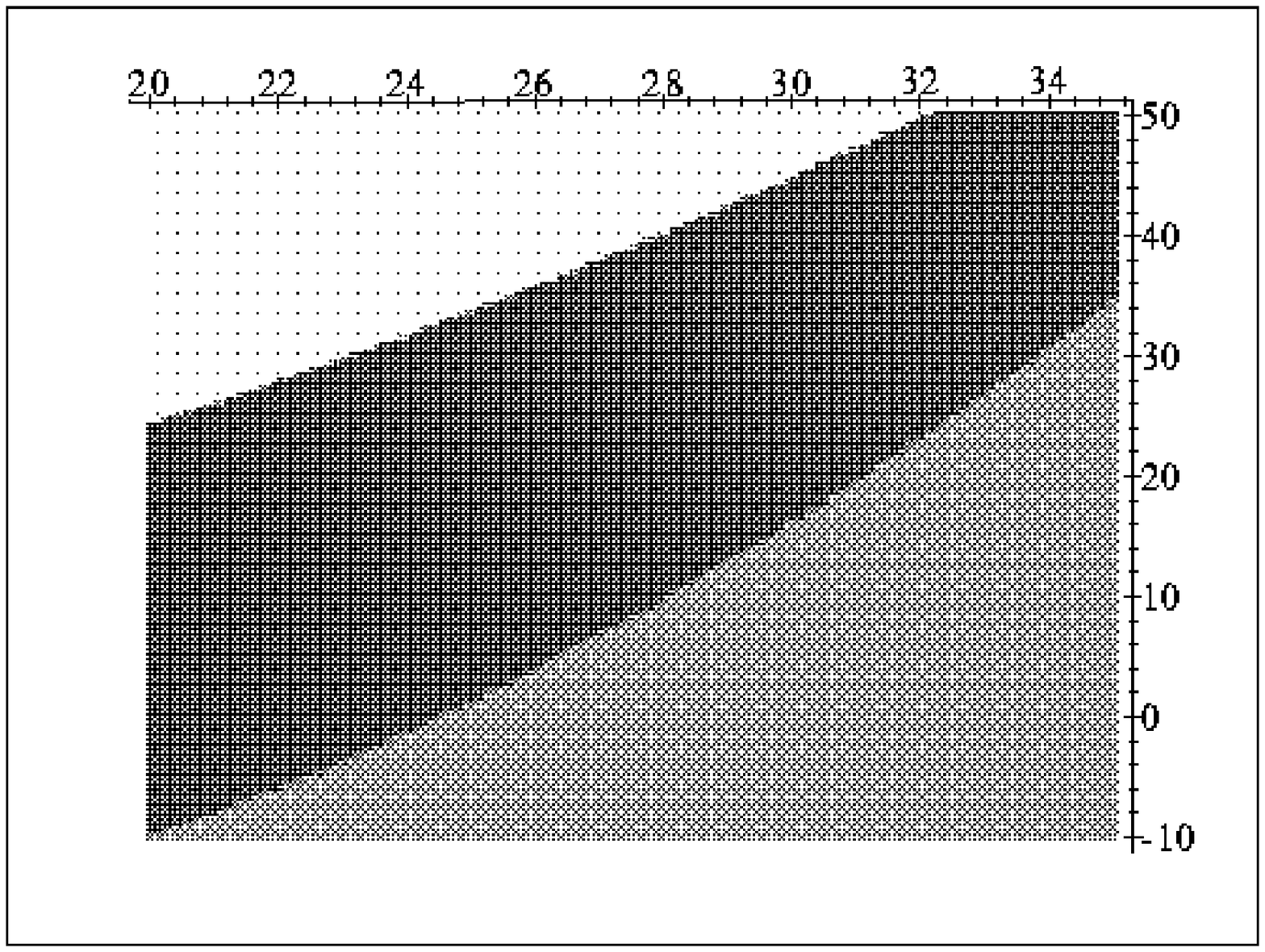,height=4.5cm,width=4.5cm,angle=-90}}
\subfigure[$r=0.9$]{\epsfig{file=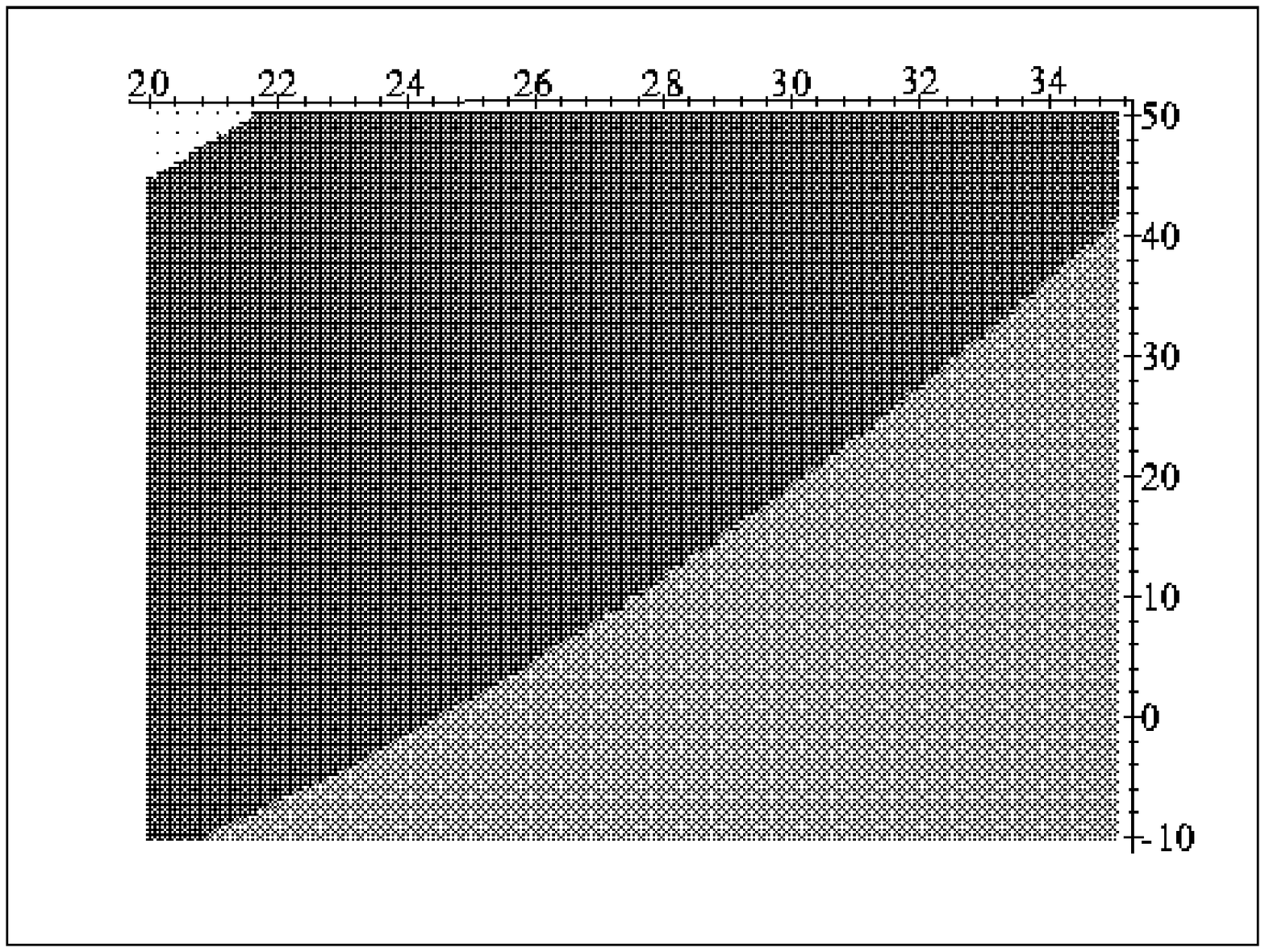,height=4.5cm,width=4.5cm,angle=-90}}
\caption{Domains of control for the third feedback scheme and several values of $r$. Vertical axis: $\eta$ from $-10$ to $50$. 
Horizontal axis: $E$ from $20$ V to $35$ V. 
Black $=0$ (stable), Grey $=1$ (unstable), White $=2$ (unstable).}
\label{dc3}
\end{center}
\end{figure}

We have numerically evaluated the index functions $g(\mu^{-1})$ of the previous Section. Figure \ref{dc12}  shows the domains
of control for the two first feedback schemes and $r=0.0$, $r=0.6$ and $r=0.9$, and 
Figure \ref{dc3} corresponds to the third feedback scheme and the same values of $r$. The black regions are those yielding index
 $0$, when the
time-delayed feedback succesfully stabilizes the $T$-periodic orbit. Several features must be highlighted:
\begin{enumerate}
\item
There are no big differences between the first and second schemes, apart from the sign of the feedback. The domain of
control of the first scheme is slightly broader. 
\item
Use of extended time-delay feedback does not improve the domain of control for the two first schemes. The index $1$ zone
expands with $r$ at the expense of both the index $0$ and index $2$ zones so, in fact, the stable region diminishes.
\item 
The third scheme is clearly inferior, as the stabilizing region is bounded and for low values of $r$ does not reach into
the chaotic region. This is not surprising, since in this scheme the feedback signal only works during one of the topologies.
In this scheme, use of extended time-delay feedback broadens the stabilizing region and allows control in the chaotic regime.
\end{enumerate}

\section{Stabilization of higher order orbits}
\label{buck_higher} 
In this Section we will show how to generalize the analytical results of Section \ref{buck_dc} to more general situations 
than period-$1$ UPOs with a single crossing per period. We will only work wirh the first feedback scheme.

Consider for instance a period-$n$ orbit with a single crossing per period at times $t_1\in(0,T)$, $(t_2\in(0,2T)$,$\ldots$,
$t_n\in((n-1)T,nT)$. Now we have to integrate $U(t)$ over $[0,nT]$ and the system changes its topology at $T, 2T,\ldots, (n-1)T$ in
addition to $t_1, t_2,\ldots,t_n$. However, both $x^*(t)$ and $x^*(t)+y(t)$ change the topology of the system at exactly 
$T, 2T,\ldots,(n-1)T$, so we may forget about these changes since they do not take the orbits apart. Mathematically, using the notation
of Section \ref{d_of_c}, this is reflected by the fact that only $\delta b(t,[x])/\delta v(t')$ enters the computations and that this
functional derivative is zero for all the variations of $b(t,[x])$ which do not depend on $v$, such as those occurrying at the end of
period.

In the first feedback scheme, we may now proceed to equations (\ref{syst}), which we will have to integrate between $0$ and $nT$ for the
two sets of initial conditions $(1,0)$ and $(0,1)$. For each cycle of the auxiliar ramp $r(t)$ we have a zero of the 
delta function and hence
$$
\delta(v^*(t)-r(t))=\sum_{k=1}^n \beta_k \delta(t-t_k),
$$
where
$$
\beta_k=\frac{1}{|\dot v^*(t_k)-\dot r(t_k)|}.
$$
Integration is now easily performed by means of a Laplace transform and we get the result
\begin{eqnarray*}
x(t) &=& \left(-x(0)\frac{a_1-\gamma_1}{2\gamma_1}+y(0)\frac{1}{\gamma_1 C}\right)e^{-\frac{1}{2}(a_1-\gamma_1)t} 
     + \left(x(0)\frac{a_1+\gamma_1}{2\gamma_1}-y(0)\frac{1}{\gamma_1 C}\right)e^{-\frac{1}{2}(a_1+\gamma_1)t} \\
     &-& \frac{E}{LC\gamma_1}\sum_{k=1}^n \left( \beta_k x(t_k) \theta(t-t_k)
     \left(e^{-\frac{1}{2}(a_1-\gamma_1)(t-t_k)}-e^{-\frac{1}{2}(a_1+\gamma_1)(t-t_k)}\right)\right)\\
y(t) &=& \left(-x(0)\frac{1}{\gamma_1 L}+y(0)\frac{a_1+\gamma_1}{2\gamma_1}\right) e^{-\frac{1}{2}(a_1-\gamma_1)t}      
     + \left(x(0)\frac{1}{\gamma_1 L}-y(0)\frac{a_1-\gamma_1}{2\gamma_1}\right) e^{-\frac{1}{2}(a_1+\gamma_1)t} \\ 
     &-& \frac{E}{2L\gamma_1}\sum_{k=1}^n \left( \beta_k x(t_k) \theta(t-t_k)
     \left((a_1+\gamma_1) e^{-\frac{1}{2}(a_1-\gamma_1)(t-t_k)}-(a_1-\gamma_1) e^{-\frac{1}{2}(a_1+\gamma_1)(t-t_k)}\right)\right)
\end{eqnarray*}
with the $x(t_k)$ computed recursively from the expression for $x(t)$.

\begin{figure}[htb]
\begin{center}
\subfigure[$r=0.0$]{\epsfig{file=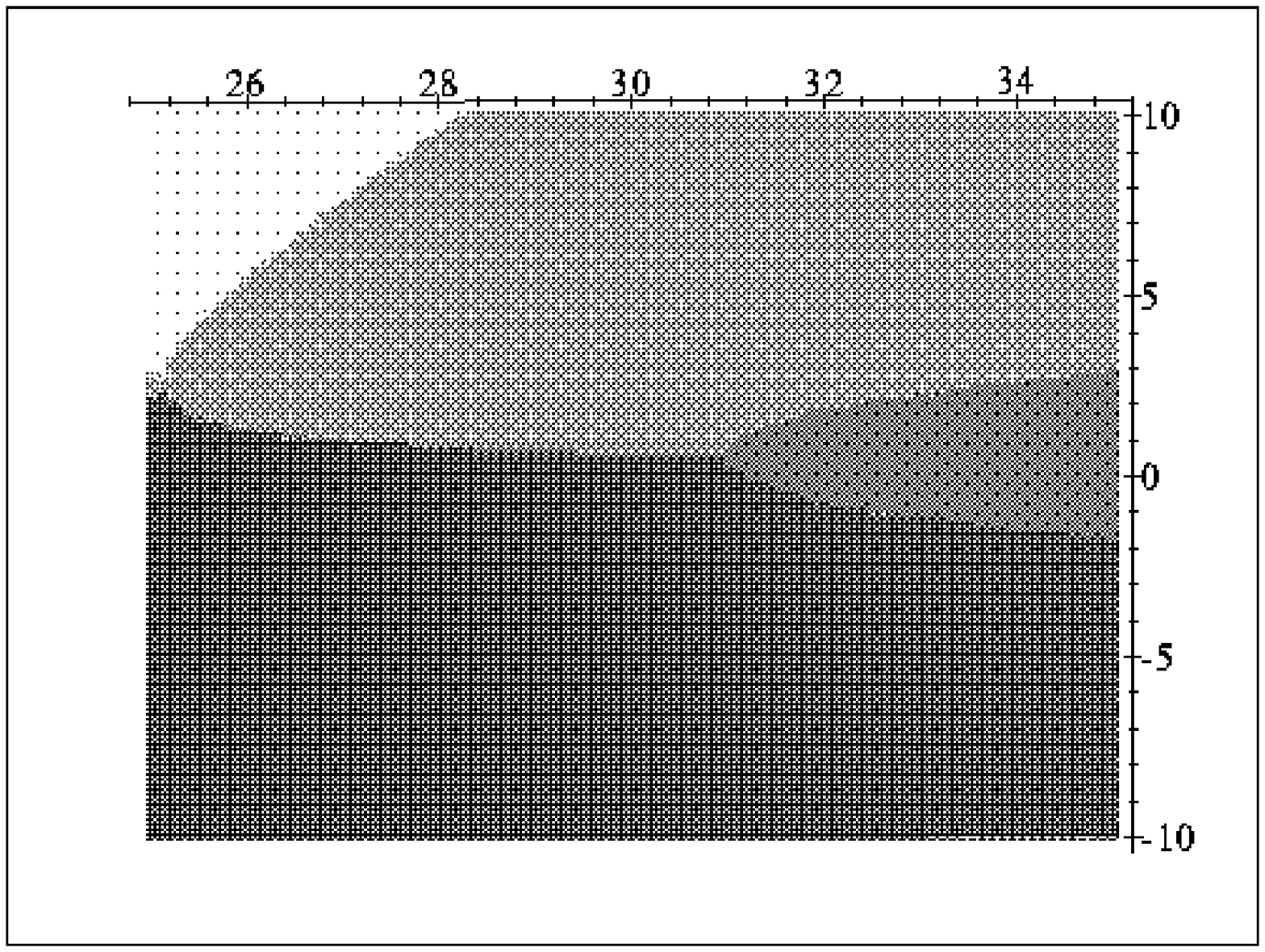,height=5.0cm,width=5.0cm,angle=-90}}
\subfigure[$r=0.6$]{\epsfig{file=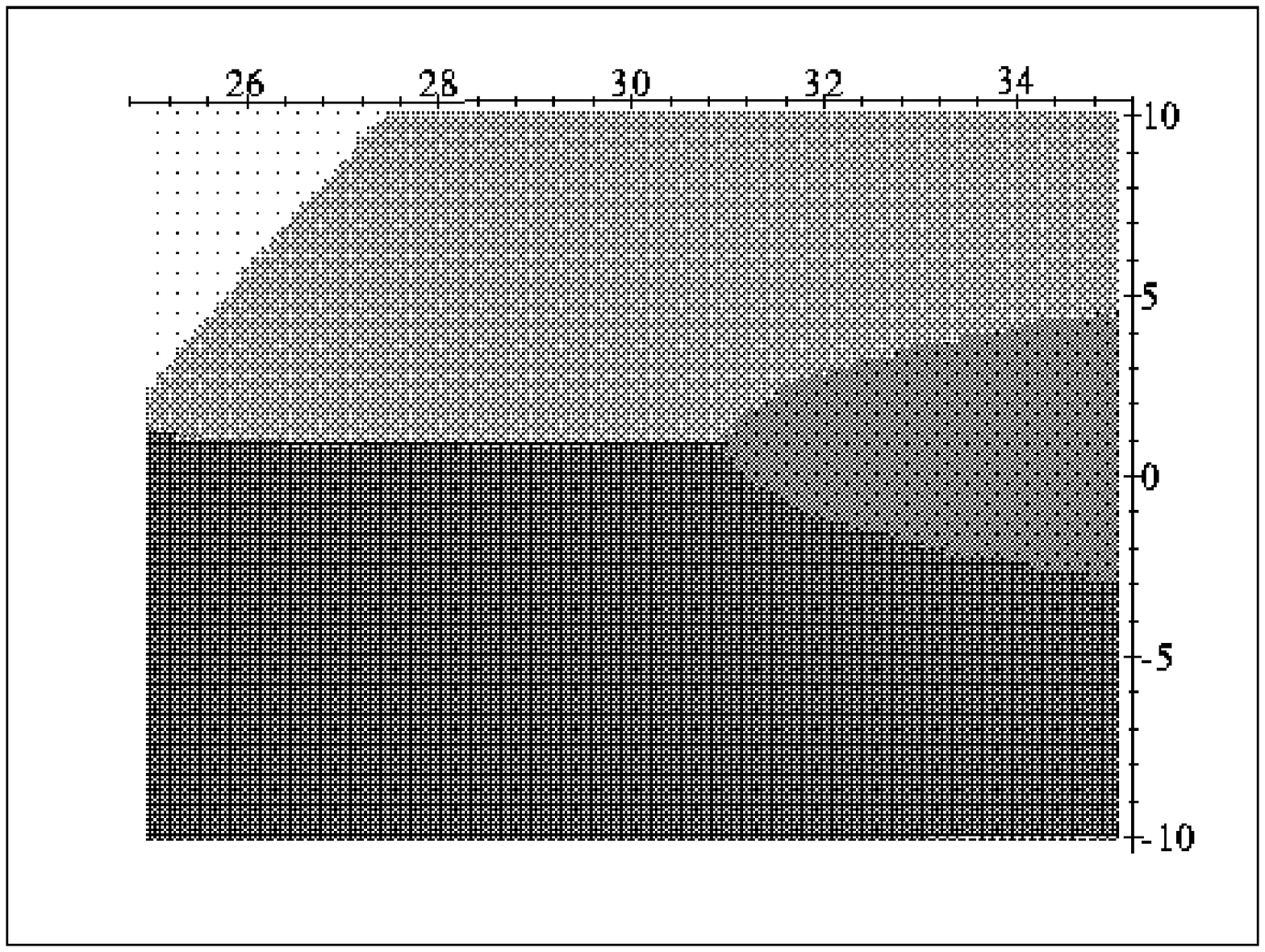,height=5.0cm,width=5.0cm,angle=-90}}
\subfigure[$r=0.9$]{\epsfig{file=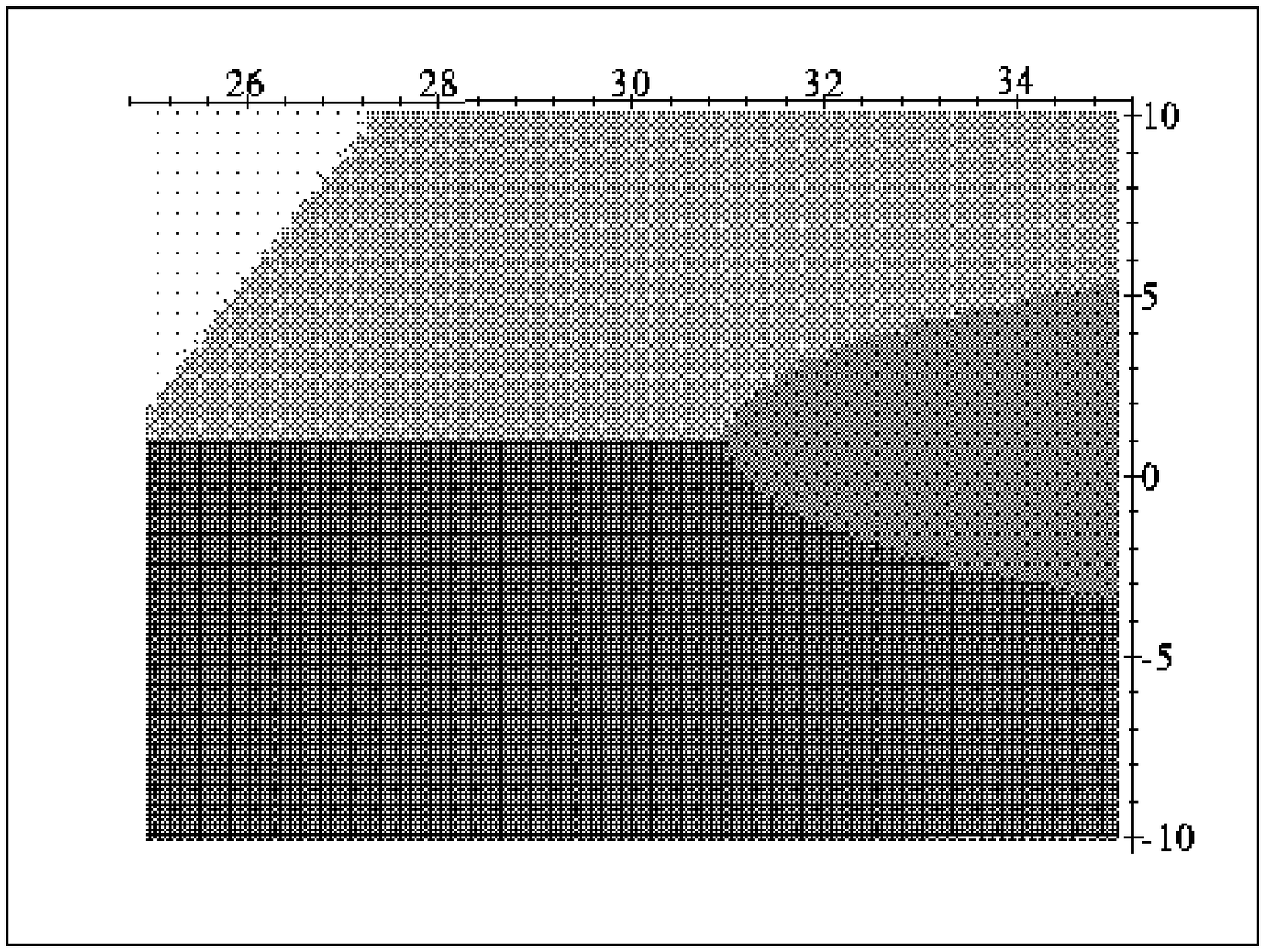,height=5.0cm,width=5.0cm,angle=-90}}
\caption{Domains of control for the first feedback scheme and period-$2$ orbits and several values of $r$. 
Vertical axis: $\eta$ from $-10$ to $10$. 
Horizontal axis: $E$ from $25$ V to $35$ V. 
Black $=0$ (stable), Dark gray $=1$ (unstable), Light gray $=2$ (unstable), White $=3$ (unstable).}
\label{dc2T}
\end{center}
\end{figure}

We will present explicit expressions for $g(\mu^{-1})$ only for the case $n=2$. One gets
\begin{eqnarray}
g(\mu^{-1})&=& \mu^{-2} e^{-2a_1 T}  \nonumber \\
   &-& 2\mu^{-1}e^{-a_1 T}\left( \cosh(\gamma_1 T)-\frac{E}{\gamma_1 L C}(\beta_1+\beta_2)\sinh(\gamma_1 T) \right.\nonumber \\
   &+& \left.   \frac{E^2\beta_1\beta_2}{\gamma_1^2L^2C^2}(\cosh(\gamma_1 T)-\cosh(\gamma_1(t_1-t_2+T)))\right) + 1.
\label{g2T}
\end{eqnarray}
Domains of control for this case with $r=0.0$, $r=0.6$ and $r=0.9$ are presented in Figure \ref{dc2T}, and the same comments
done for the period-$1$ orbits do apply here. Other situations, such as period-$1$ orbits with several crossings per period,
 can be treated along the same lines  developed in this Section.    

\section{ETDAS simulations}
\label{buck_sim}
In this Section we report the results of several simulations of the time-delay controlled system.

\begin{figure}
\begin{center}
\epsfig{file=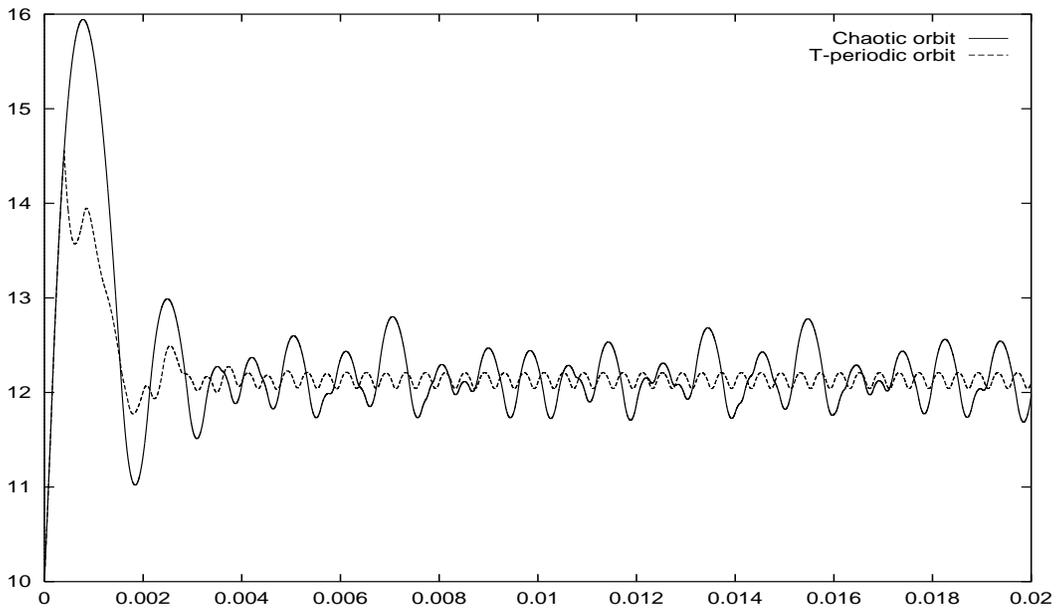,height=15cm,width=8cm,angle=270}
\caption{Load voltage chaotic waveform for $E=35 \mbox{V}$ (solid line) and time-delay feedback controlled orbit 
in the second scheme with
$r=0$ and $\eta=4.0$ (dashed line). $50$ periods of the auxiliary ramp are shown.}
\label{exemple}
\end{center}
\end{figure}

Figure \ref{exemple} shows a typical numerical simulation of the time-delay feedback control method. A chaotic orbit of the 
system ($\eta=0$) is also
shown for reference. The feedback starts to act after the first period and stabilizes the orbit in less than $10$ periods, 
\textit{i.e.}, $4\ \mbox{ms}$ for the system considered.

\begin{figure}
\begin{center}
\subfigure[]{\epsfig{file=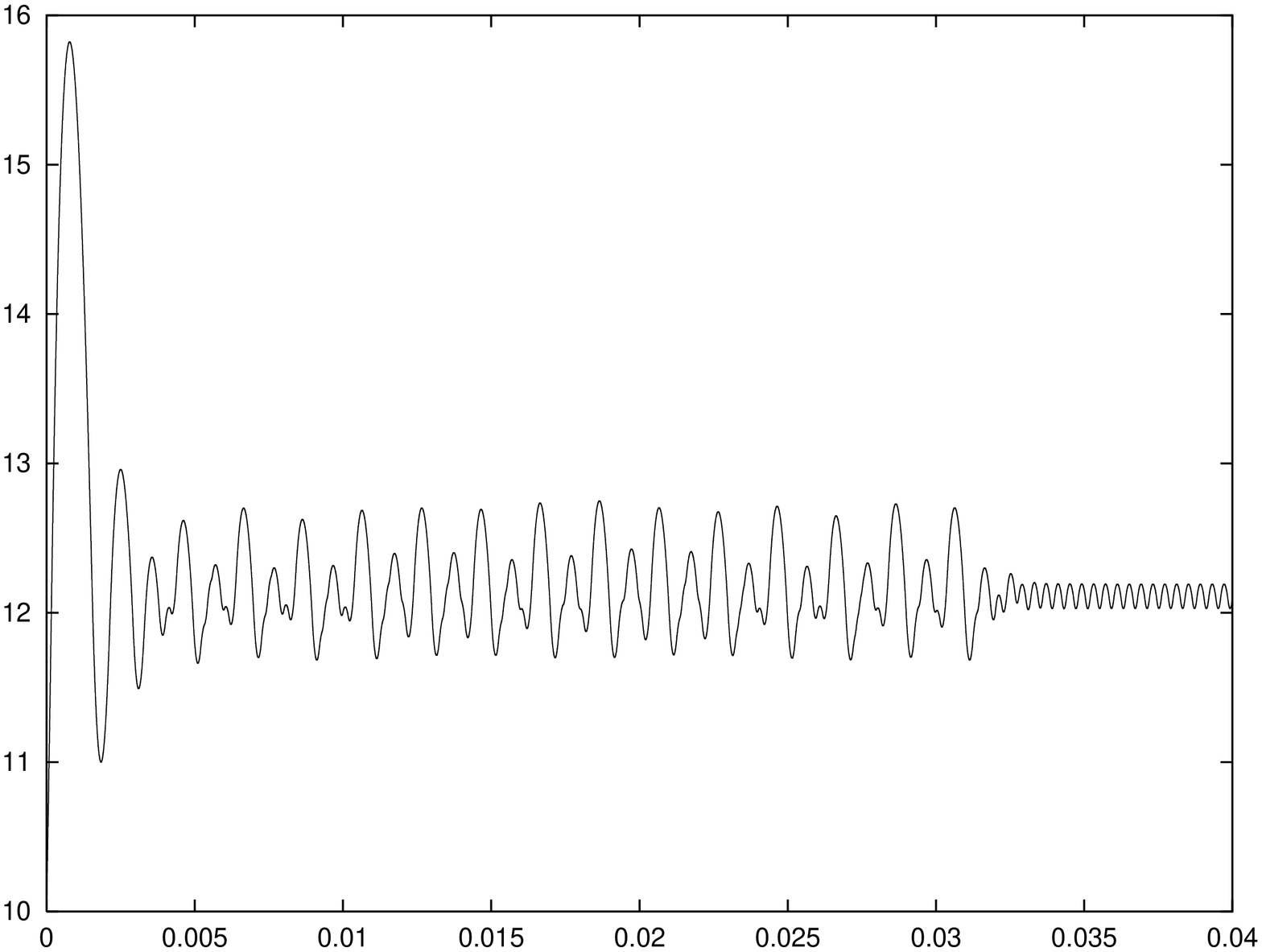,height=7.5cm,width=7.5cm,angle=270}}
\subfigure[]{\epsfig{file=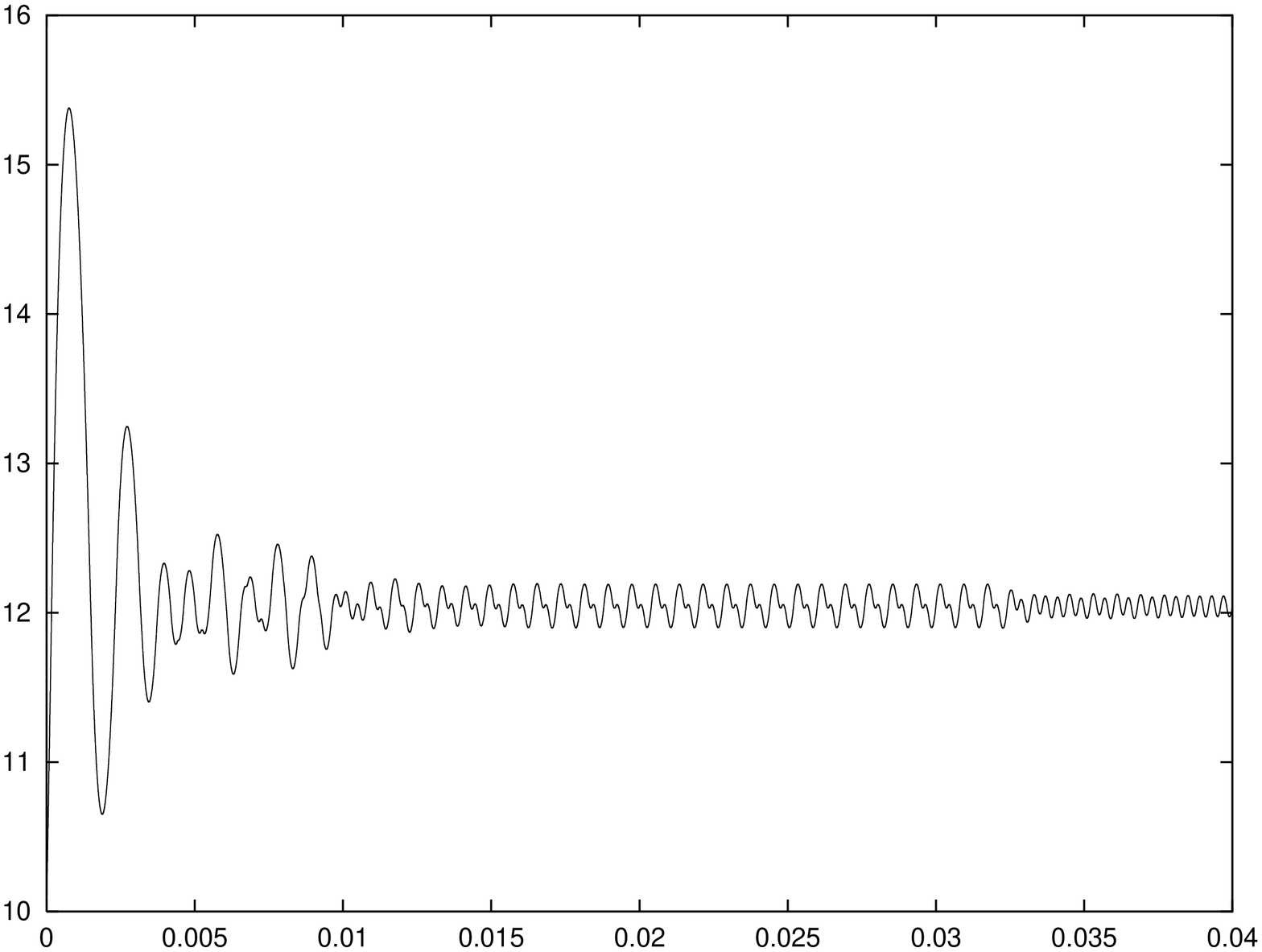,height=7.5cm,width=7.5cm,angle=270}}
\caption{ETDAS simulations for the \textit{buck} converter}
\label{ETDASexemple}
\end{center}
\end{figure}

Figure \ref{ETDASexemple}(a) shows a simulation of an ETDAS with $r=0.6$, $\eta=-5.0$ and $E=33 \mbox{V}$ using the first
scheme. The series in (\ref{eq2}) has been truncated to the first $80$ terms. The figure shows the first $100$ periods, and
the feedback starts to act after $t=80\cdot T$. Notice that the collapse of the chaotic regime to the stabilized period-1
orbit is nearly instantaneous.Another ETDAS, this time using the third scheme and $r=0.6$, $\eta=6.0$ and $E=26 \mbox{V}$ is
depicted in Figure \ref{ETDASexemple}(b). The system first stabilizes on the period-2 stable orbit. The time-delay control starts to act
again after $t=0.032 \mbox{s}$ and, after a short transient, the period-1 UPO is stabilized.

In order to check the frontier between the zones of the domain of control, we have numerically integrated
the time-delay feedback equations for several values of the bifurcation parameter $E$ and 
the feedback gain $\eta$ on both sides of the analitically computed frontier, although some times numerical integration errors
may produce a wrong result if $\eta$ is too close to the frontier (this is particularly true of the third feedback scheme, where
numerical errors during the $v(t)<r(t)$ topology are amplified during the phase $v(t)>r(t)$, when the control does not act).

\begin{figure}
\begin{center}
\subfigure[]{\epsfig{file=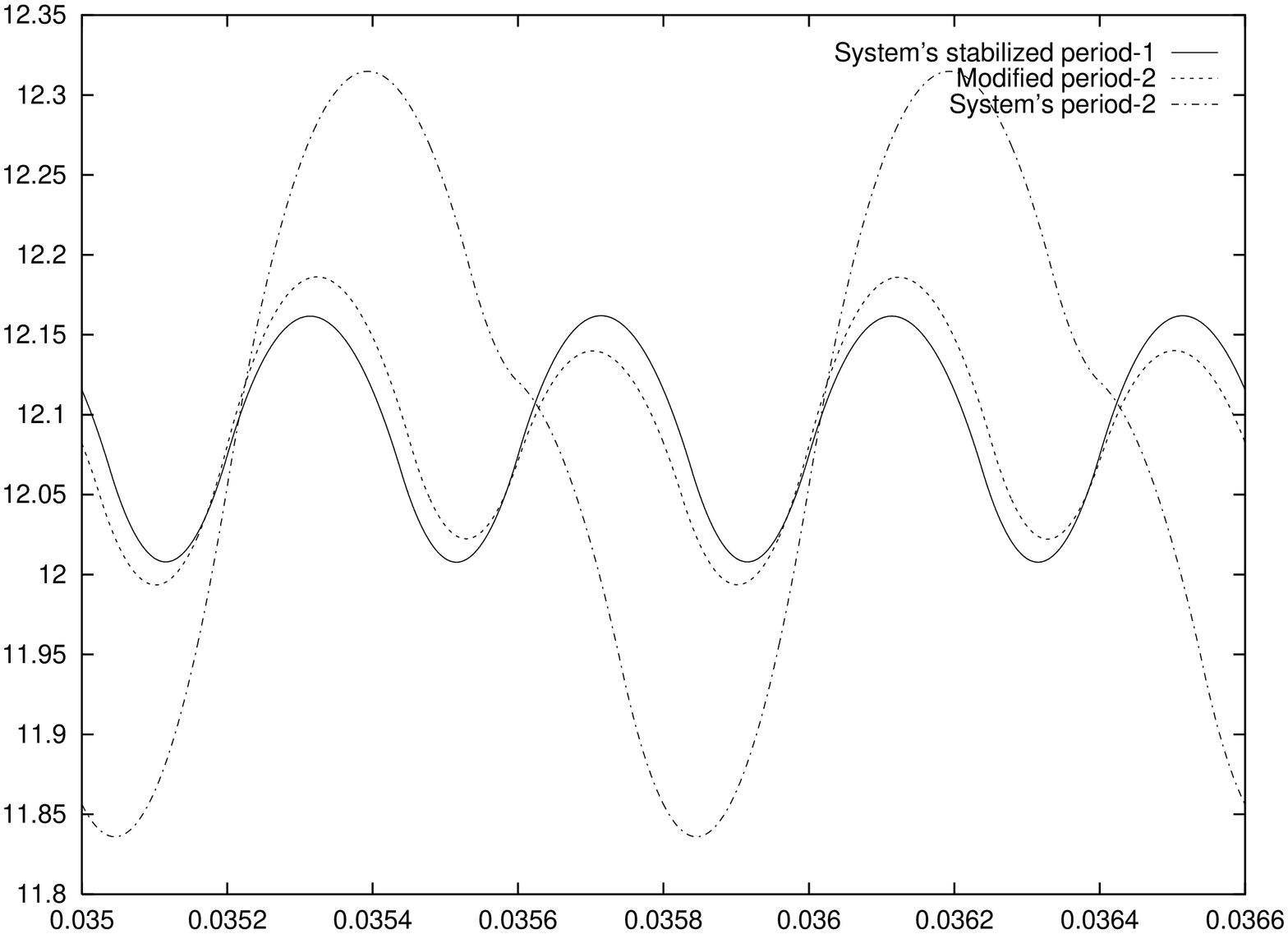,height=7.5cm,width=7.5cm,angle=270}}
\subfigure[]{\epsfig{file=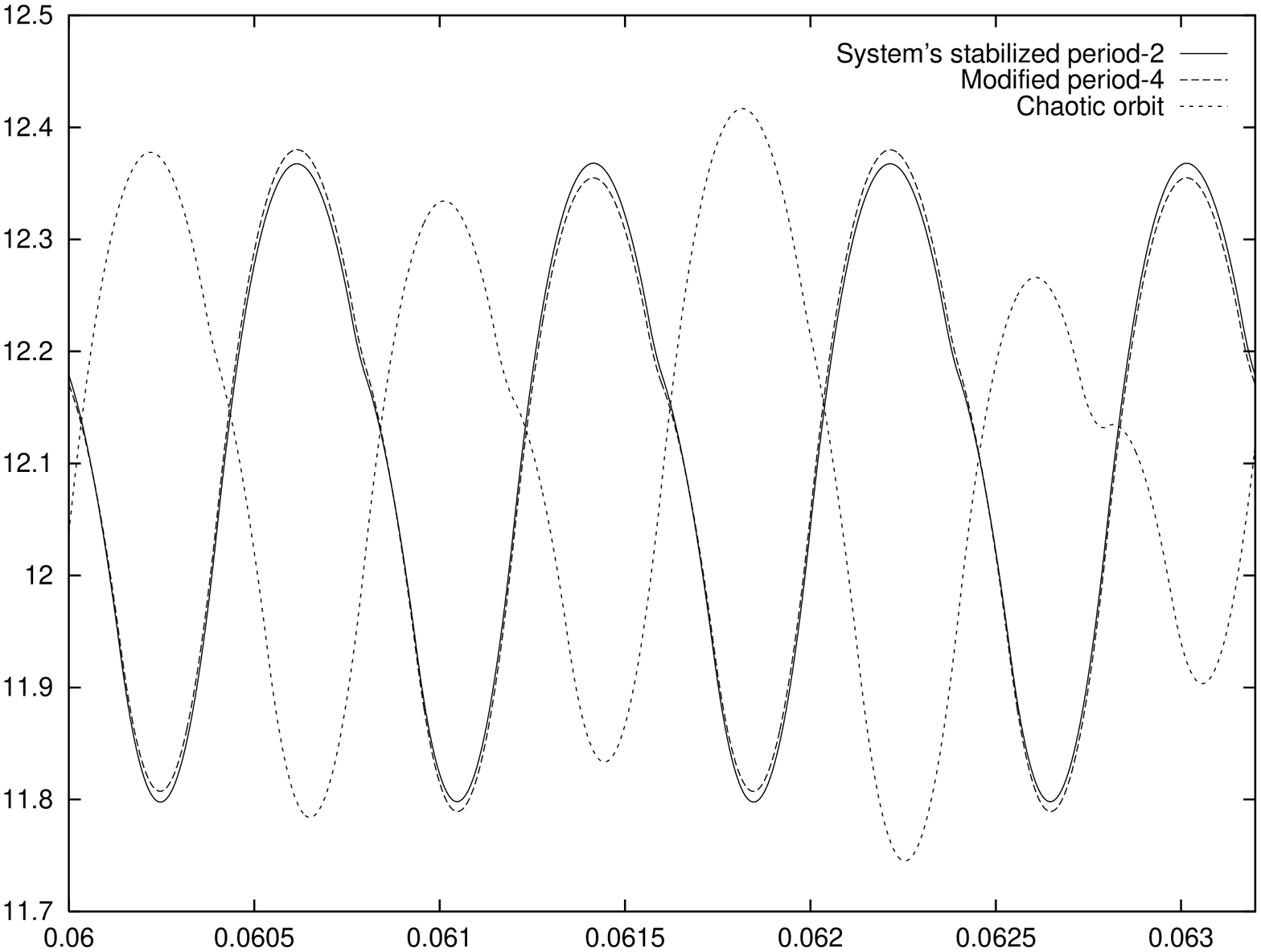,height=7.5cm,width=7.5cm,angle=270} }
\caption{Simulation checks of the analytically computed limits of the control domains}
\label{sim1}
\end{center}
\end{figure}

One of those checks is presented in Figure \ref{sim1}(a), which corresponds to input voltage $E=30$ with 
the first feedback scheme and $r=0$. The solid line corresponds
to $\eta=-1.3$ and the dashed one to $\eta=-1.2$. The expression (\ref{eqs2}) for $g_1(\mu^{-1})$ predicts index $0$ for the former
and index $1$ for the later. The time span in the figure corresponds to four periods of the auxiliary ramp 
and the vertical axis represents the capacitor voltage. 
We see indeed that $\eta=-1.3$ stabilizes the system to the period-1 orbit, while $\eta=-1.2$ does not. In fact,
$\eta=-1.2$ produces a period-2 orbit  which, however, is 
not the stable period-2 orbit of the uncontrolled system which exists
for this value of $E$ and is also represented in the figure: there is some kind of competition between the stable period-2 orbit and the time-delay feedback (which is
not zero on this period-2 orbit) which finally neither stabilizes to the unstable period-1 orbit nor 
falls on the system's stable period-2 orbit.

Another check, this time for period-2 orbits, is illustrated by Figure \ref{sim1}(b), corresponding to $E=32.5$ and $r=0.0$. 
Eight periods of the auxiliar ramp, between $t=0.06$ and $t=0.0632$, are represented.
Equation (\ref{g2T}) predicts index $0$ for $\eta=-1.1$ and index $1$ for $\eta=-1.0$. A chaotic orbit of the system
is also represented, and the same comment done for Figure \ref{sim1} applies here. Notice that for this value of $E$ there is also
a period-1 UPO, which, however, is not stabilized by the above values of $\eta$. However, for $\eta\lesssim -1.6$, the system
could choose to stabilize on either the period-1 or period-2 UPOs, since both are zero on the feedback control in this case. At
present we do not have a way to predict which orbit would the system choose in a given case, or whether there are well defined
basins of attraction.

\section{Conclusions}
\label{conc}
We have obtained explicit, analytically computed expressions for the function  $g: S^1\rightarrow \C$ whose index around the origin
yields the domain of control of the ETDAS method for a variety of orbits and feedback schemes of the \textit{buck} converter. 
The analytical computation is possible due to the linear character of the differential equations of both topologies of the \textit{buck}
converter and the fact that the change of topology only introduces a delta function in the relevant equations, which can be translated
to a delta function in the time variable. Things are not so easy for other DC-DC converters. For instance, for the current-controlled
boost, one gets a delta function of the time derivative of the intensity, which is not differentiable at the topology-changing points.

Of the three feedback schemes that we have presented, the two first are clearly superior in the sense that the feedback gain needed
to stabilize the UPOs increases slowly as we enter the chaotic zone, while the third scheme is unable to 
stabilize the UPOs in the chaotic regime for low values of $r$ and, in any case, higher values of the feedback gain
than in the other two cases are needed. The domain of control is quite simple in all the cases, and does not
change dramatically as large values of $r$ are used, although extended delay enlarges the stabilizing region in the third scheme.
The simplicity of the domain of control for the \textit{buck} converter makes it plausible that
its general features will not change in a relevant way if nonideal elements, such as resistors in the inductor or in the voltage source
that delivers $\Delta v$ are introduced, or if small noise is considered.

We have numerically confirmed our results by simulating the time-delayed system and checking the predicted behaviour when 
varying the feedback gain. The simulations show that ETDAS is quite effective for the \textit{buck} converter, although presently
we don't have a way to force the system to choose a given orbit when several ones compatible with the imposed period are available.

Work is in progress to extend our results to other DC-DC converters and to include the effects of nonideal elements and noise.

\section*{Acknowledgements}
This work has been partially supported by CICYT under Project TAP94-0552.

\end{document}